\documentclass{aa}  

\usepackage{graphics}
\usepackage{txfonts}
\usepackage{soul}
\usepackage[utf8]{inputenc}
\usepackage{amsmath, amssymb}
\usepackage{subfig}
\usepackage{xcolor}
\usepackage{float}
\usepackage{makecell}
\usepackage{multirow, array}
\usepackage[normalem]{ulem}
\usepackage[switch]{lineno}
\usepackage[pdfencoding=auto, psdextra]{hyperref}
\hypersetup{
    colorlinks=true,
    linkcolor=blue,
    filecolor=magenta,      
    urlcolor=cyan,
    }

\newcommand{\rev}[1]{{\color{black}#1}} 
\begin{document}

   \title{BGRem: A background noise remover for astronomical images based on a diffusion model}

   \subtitle{}

   \author{Rodney Nicolaas\inst{\ref{inst1}}\thanks{Corresponding Author: \texttt{rodney.nicolaas@ru.nl}} 
   \and Sascha Caron\inst{\ref{inst1}}\inst{,\ref{inst2}} 
   \and Fiorenzo Stoppa\inst{\ref{inst1}}\inst{,\ref{inst3}} 
   \and Saptashwa Bhattacharyya\inst{\ref{inst4}}\thanks{Corresponding Author: \texttt{sbhattacharyya@ung.si}} 
   \and Roberto R. de Austri\inst{\ref{inst5}} 
   \and Paul J. Groot\inst{\ref{inst1}}\inst{,\ref{inst6}}\inst{,\ref{inst7}}
   \and Andrew J. Levan \inst{\ref{inst1}}
   }

   \institute{Institute for Mathematics, Astrophysics and Particle Physics (IMAPP), Radboud University Nijmegen, Heyendaalseweg 135, 6525 AJ Nijmegen\label{inst1} 
   \and  Nikhef, Science Park 105, 1098 XG Amsterdam, the Netherlands\label{inst2}
   \and  Department of Physics, University of Oxford, Denys Wilkinson Building, Keble Road, Oxford OX1 3RH, UK\label{inst3} 
   \and University of Nova Gorica, Centre for Astrophysics and Cosmology, Vipavska 11c, Ajdovščina, Slovenia\label{inst4} 
   \and Instituto de Física Corpuscular (IFIC), CSIC‐UV, Spain\label{inst5} 
   \and Department of Astronomy and Inter-University Institute for Data Intensive Astronomy, University of Cape Town, Private Bag X3, Rondebosch 7701, South Africa\label{inst6} 
   \and South African Astronomical Observatory, PO Box 9, Observatory, Cape Town 7935, South Africa\label{inst7}}

  \date{Received Month day, year; accepted Month day, year}

% \abstract{}{}{}{}{} 
% 5 {} token are mandatory

% Paul Groot suggested changing all figure X into Figure X
 
\abstract{Context: Astronomical imaging aims to maximize signal capture while minimizing noise. Enhancing the signal-to-noise ratio directly on detectors is difficult and expensive, leading to extensive research in advanced post-processing techniques.

Aims: Removing background noise from images is a valuable pre-processing step catalog-building tasks. We introduce BGRem, a machine learning (ML) based tool to remove background noise from astronomical images. Our aim is to improve image quality and enhance the performance of the subsequent analysis pipeline, from detecting faint sources to source characterization tasks.
% and machine learning models, enabling more accurate and reliable astronomical observations.}

Methods: BGRem uses a diffusion-based model with an attention U-Net as backbone, trained on simulated images for optical and gamma $(\gamma)$-ray data from the MeerLICHT and Fermi-LAT telescopes. In a supervised manner, BGRem learns to denoise astronomical images over several diffusion steps. Pre- and post-processing techniques, including normalization and median subtraction, are performed on these images to make them suitable for the analysis pipeline.

Results: BGRem performance was compared with a widely used tool for cataloging astronomical sources, SourceExtractor (SExtractor). It was shown that the amount of true positive sources using SExtractor increased by about $7\%$ for MeerLICHT data when BGRem was used as a pre-processing step. We also show the generalizability of BGRem by testing it with optical images from different telescopes and also on simulated $\gamma$-ray data representative of the Fermi-LAT telescope. We show that in both cases, BGRem improves the source detection efficiency.

Conclusions: BGRem can improve the accuracy in source detection of traditional pixel-based methods by removing complex background noise. Using zero-shot approach, BGRem can generalize well to a wide range of optical images. The successful application of BGRem to simulated $\gamma$-ray images, alongside optical data, demonstrates its adaptability to distinct noise characteristics and observational domains. This cross-wavelength performance highlights its potential as a general-purpose background removal framework for multi-wavelength astronomical surveys.}


\maketitle

%%%%%%%%%%%%%%%%%%%%%%%%%
%\input{Introduction}
% Introduction
%%%%%%%%%%%%%%%%%%%%%%%%%

\section{Introduction}
Astronomical imaging is fundamentally challenged by the need to maximize signal capture while minimizing noise interference. Achieving a high signal-to-noise ratio (SNR) is crucial for ensuring the reliability and accuracy of scientific analyses derived from these images. However, enhancing SNR directly on the detector is both technically and financially difficult.  While the common form of noise is shot noise (Poisson noise), related to random fluctuations in the photon counts, the instrumental noises like readout noise and dark current noise also affect the image quality. While increasing the exposure time can increase the signal, the noise will increase too. It is also possible to stack images, known as `coaddition', to increase the SNR and depth of the images. However, these images could be taken under different atmospheric conditions (transparency), background, and some instruments have different optical quality (Point Spread Function, PSF) across different parts of the sky. Methods to improve the process of coaddition are an active area of research \citep{Zackay-Coadd1, Zackay-Coadd2}, as it plays a significant role in the ongoing surveys like ZTF \citep{Bellm-ZTF}, and in the near future, LSST \citep{Ivezic-LSST}. Several traditional methods, such as bilateral filtering, wavelets-based method, non-local means, block matching, etc., (for a comprehensive review, check \cite{denoise-review}) are often used as processing steps for astronomical image enhancement tasks.  

Over the last decade, several ML-based algorithms, specifically Deep Neural Network (DNN) inspired techniques, have been used for image denoising tasks (for a review of modern techniques, check \cite{TIAN-DL-denoise-Review}). ML models are highly dependent on the quality and characteristics of their training data. Many models are trained on simulated data where the ground truth is controlled, and assumptions about background noise are predefined. However, generalizing these models to real data, which always includes noise, remains a significant challenge. For astronomical images, a deep convolutional autoencoder network was shown to be effective for denoising radio images and detecting faint sources \citep{radio-deepAutoencoder}. For optical images coming from the Hubble Space Telescope, U-Net-based \citep{u-net} denoising network, `Astro U-Net' \citep{u-net-hst} was shown to produce images with noise characteristics similar to those obtained with twice the exposure. Recently, for large-scale sky surveys, a DNN-based image restoration algorithm with the active learning strategy has been tested \citep{active-learning-PSF}, and shown to minimize the impact of noise and PSFs.

Going beyond image denoising tasks, \cite{wolf2024direct} showed that their convolutional autoencoder neural network increased sensitivity for direct imaging of exoplanets by up to 2.6 times compared to conventional methods. \cite{ehlert2022probabilistic} developed a probabilistic method to remove the background in low signal X-ray and $\gamma$-ray event list data gathered by the observatories. They showed that using their method, which takes into account the random Poisson fluctuations, the detection of galaxy clusters can be more significant than with standard methods of background subtraction. \cite{acciarri2021cosmic} utilized a UResNet to remove noise from neutrino detections, employing segmentation to differentiate between neutrino interaction pixels and background pixels, showing promising results on simulated data. In their search for a rare decay in a particle detector experiment, \cite{tung2024suppression} successfully integrated a convolutional neural network with the statistical technique of pulse shape discrimination. This approach enabled them to suppress neutron background events by a factor of $5.6\cdot10^5$, while maintaining a photon signal acceptance rate of $\sim 70\%$.

In this work, we present BGRem, an ML–based tool, built on the Gaussian diffusion framework \citep{sohl2015deep, jon-ho-diffusion, yang-score-diff}, for removing background noise from astronomical images. Diffusion models form a category of generative models and are effective for text-to-image generation, image inpainting, superresolution, and other applications (for a review of diffusion models in vision, check \cite{Diffusion-Review}).  Here, in BGRem, we use inspiration from the diffusion model for denoising tasks under a supervised ML scheme with an attention U‑Net \citep{oktay2018attention} as the backbone. To quantify and compare our results, we used the background subtraction method in SExtractor \citep{bertin1996sextractor} for astronomical images. SExtractor automatically detects sources in astronomical images, with background subtraction being a key pre-processing step. Our experiments show that BGRem removes background noise for optical images from different telescopes as well as $\gamma$-ray simulated data, with high accuracy within a reasonable computation time compared to SExtractor. 

This study is an addition to a set of works to build an end-to-end ML-based pipeline, ASID \citep{panes2021identification}, to enhance and complement traditional pixel-based methods. Previously, we have developed astrophysical source detection and localization pipelines for optical \citep{stoppa2022autosourceid-L}), and gamma-ray sources \citep{panes2021identification, chris-gamma}, which improve upon pixel-based and likelihood-based methods. For further characterization of the detected sources using Deep Neural Nets (DNNs), e.g. flux estimation and star-galaxy classification,  ASID-FE \citep{stoppa2023autosourceid-FE} and ASID-C \citep{stoppa2023autosourceid-c} were developed. We show the potential of BGRem to be included as a background removal step before applying the source detection and localization pipeline to further enhance the catalog-building task. All the codes used for the production of the results shown here are publicly available at GitHub\footnote{\href{https://github.com/RNicolaas/BGRem}{GitHub Link} \includegraphics[height=4.0mm]{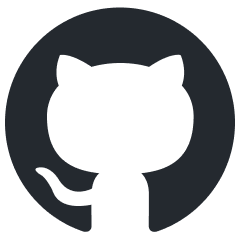}}. 

In Section \ref{sec:training-data}, we elaborate on the training data and pre-processing methods employed in our study. Section \ref{sec:model} delves into the intricacies of the model, including post-processing methods. The results are presented in Section \ref{sec:results}, where we evaluate the performance of BGRem on both test datasets and real-world applications. Finally, Section \ref{sec:conclusion} offers a summary of our findings and explores future research directions.

%%%%%%%%%%%%%%%%%%
%\input{train-data}
% Training Data
%%%%%%%%%%%%%%%%%% 

\section{Training Data Preparation}\label{sec:training-data}

Since BGRem is based on a supervised ML scheme, here we describe the training data production procedure. 

The training data consisted of 5 simulated full-field images from the MeerLICHT telescope \citep{groot2024blackgem}, an optical telescope at the University of Cape Town\footnote{https://science.uct.ac.za/meerlicht}. Three more images were simulated and used as test data. For the simulation, we have used the GalSim \citep{rowe2015galsim} simulation software \rev{with the corresponding background components: a) a spatially varying sky background, b) detector-level effects including Gaussian read noise and dark current, c) Poisson noise fluctuations coming from both the sky and the dark current.}

Using simulation is advantageous for training BGRem since it provides ground truth images, which are the source-only images, without any background noise. In contrast, such ground truth is fundamentally inaccessible from real observations; one can generate background-reduced images using a specific model or estimation algorithm, but these should not be considered as ground truth for training an ML algorithm. Simulated data give us the possibility of a controlled supervised training, but they may not fully capture the complexity of the real observational artefacts, potentially affecting model generalization and driving the interpretation of the data as shown before for $\gamma$-ray data \citep{Caron2023}. Later in the Results section (sec: \ref{sec:results}), we discuss the application and performance of BGRem on real data.

Similar to the actual MeerLICHT data, the simulated images have a shape of $10560 \times 10560$ pixels, which would be too large and memory-consuming as an input to a DNN for processing. Therefore, we cut them into $256\times 256$ pixel sub-images, resulting in a total of $8405$ images. We used $10\%$ of these images as test data, leaving approximately $7500$ images for training. To ensure the quality of the training data, we excluded images whose maximum pixel value was below 500. This threshold, a hyperparameter of the training process, primarily affects training time and the models' performance on brighter sources. This threshold was chosen to maximize the amount of training data used, while removing the images without any bright sources.

Before an image is fed into the model, it undergoes two pre-processing steps. First, the median pixel value is subtracted. Since most of the image consists of background noise, the median pixel value provides an estimate of this background. By subtracting this, the noise will be centered around zero. This works only if the background is flat across the entire image, which is generally not the case. Therefore, an alternative method can be used to divide the image into smaller images and do the same for each smaller image individually. The second step is normalization, ensuring that the pixel values fall within a range similar to the training data. This normalization factor can be set manually or estimated by BGRem. It is crucial that the input background noise has the same spread as the training data, allowing the model to recognize the noise in every image. 

To study the possibility of extending BGRem at a wavelength different from optical, and for different background noise, we use simulated gamma-ray ($\gamma$-ray) sky images. For this purpose, we used the procedure described in \cite{panes2021identification} for generating simulated $\gamma$-ray sky maps for the Fermi-LAT \citep{Atwood2009} observation within $1-2$ GeV range. The simulated data is based on 10 years of observation of LAT, and it includes the source properties of two of the most common source classes, Active Galactic Nuclei (AGNs) and Pulsars (PSRs). One of the biggest challenges in analyzing the $\gamma$-ray sky for detecting and characterizing these sources is the presence of the Interstellar Emission (IEM) \citep{Acero-IEM, 4fgl2020}, which results from the interaction of the energetic Cosmic Rays with the Interstellar Medium. Specifically, close to the Galactic plane $(|b|<20^{\circ})$, where the contribution from the IEM is significant, could result in spurious source detection and also hide a faint source population. To test the possibility of removing this IEM with BGRem, we have used the latest Fermi-LAT background model\footnote{https://fermi.gsfc.nasa.gov/ssc/data/access/lat/\\BackgroundModels.html} \citep{4fgl2020} to generate $\gamma$-ray sky patches including point sources and IEM.  To train the BGRem to learn and remove this IEM component from the images, we have used 12000 patches of size $64\times 64$ distributed randomly over the whole sky and tested the performance with 600 patches.  These images are essentially photon counts representing an infinite statistics scenario, and for the 1-2 GeV energy bin, considering the LAT PSF, the pixel resolution is $\sim 0.2 ^{\circ}$ (deg.). \rev{We have currently tested BGRem for 1-2 GeV data as a proof-of-concept study, motivated by practical considerations. In this range, the LAT point-spread function is relatively narrow $(\sim 0.2 ^{\circ})$ compared to lower energies, while photon statistics remain substantially higher than at higher energy bins $(>2 ~\text{GeV})$, where source counts are negligible. This energy band, therefore, provides a well-balanced regime to avoid the effects of bad PSF and low statistics. To apply BGRem at a different energy range, a thorough hyperparameter search has to be performed during training, however, the current framework can act as a starting point.}

%%%%%%%%%%%%%%%%%%%%%%
%\input{model}
% BGRem Model
%%%%%%%%%%%%%%%%%%%%%%%

\section{The Model: BGRem Framework}\label{sec:model}

The working engine behind BGRem is inspired by the diffusion model. Standard diffusion models training starts with a clean image and iteratively adds noise during the forward process via a Markov Chain until it is turned completely into Gaussian noise of zero mean and unit variance \citep{sohl2015deep, jon-ho-diffusion}. The goal of the reverse process is to learn the incremental iterative transformation starting from a Gaussian noise to produce a coherent image in an unsupervised manner.

BGRem takes astronomical images as input, which are made of the ground truth images (astrophysical sources only) with scaled noise added onto them, and can be described as: 
\begin{align}\label{eq:add-noise}
x_t = y + \sigma_t \cdot x    
\end{align}
where $x_t$ is the noisy image at diffusion step $t$, $y$ is the clean (source-only) image, $\sigma_t$ can be defined as noise scale, and $x$ is the background noise. One can think of this as an addition of a portion of the background noise $(\sigma _t \cdot x)$ to the clean image, where we sample a random $t \, (t \in \left[0, 1\right] \sim \text{Uniform (0, 1)})$ from a uniform distribution. This sampled time is mapped to a pair of signal and noise scaling factors using a cosine schedule:
\begin{align}
    \theta _t = \theta _{\text{start}} + t \cdot (\theta _{\text{end}} - \theta _{\text{start}})
\end{align}
where the angle $\theta _t \in [\theta _{\text{start}}, \, \theta _{\text{end}}]$ is an interpolation between predefined bounds corresponding to maximum $(\sim 1)$ and minimum $(\sim 0)$ signal rates.
The noise scaling term $(\sigma _t)$ is the time-dependent term that is governed by a cosine-based diffusion schedule, which smoothly interpolates between high signal and high noise regimes. Specifically, the signal $(\alpha _t)$ and noise rates $(\sigma _t)$ are defined as:
\begin{align}\label{eq:sigma-sched}
\alpha _t = \text{cos}(\theta _t); \, \, \sigma _t = \text{sin}(\theta _t)
\end{align}
 This signal rate does not directly affect the images, but this trigonometric parametrization for signal and noise rate ensures that the total variance is constant $(\alpha _t^2 + \sigma _t^2 =1)$ across all time steps and ensures numerically stable training. 

To inform the network of the current noise level, the squared noise scale $(\sigma _t^2)$ is passed through a sinusoidal positional embedding, following the standard approach in diffusion models. This embedding is then upsampled and concatenated with the input image features, allowing the attention U-Net to condition its denoising behavior on the level of corruption present at each timestep. BGRem focuses on predicting the noise component, which could then be subtracted from the noisy image to produce the denoised image as below:
\begin{align}
\hat{y} = x_t - \sigma _t \cdot \hat{x}
\end{align}

To train the model, we use the mean absolute error as a loss function, which compares between true and predicted noise as shown in Eq. \ref{eq:lossl1}, leading to a supervised diffusion denoising technique. 
\begin{align}\label{eq:lossl1}
	\rev{\mathcal{L} = \mathbb{E}_{y\sim p_{data}}\mathbb{E}_{t\sim \text{Uniform}(0, 1)}\mathbb{E}_{x\sim p_{noise}}[||x-\hat{x_{\phi}}(x_t, t) ||_{1}] 	}
\end{align}	
\rev{Here $\hat{x}_{\phi}$ denotes the network prediction parameterized by the network parameters $\phi$. Since the original Denoising Diffusion Probabilistic Model (DDPM) work \citep{jon-ho-diffusion} used Mean-Squared-Error (MSE) as a loss function, we later highlight in Appendix \ref{sec:app-mse} on potential drawbacks of training with MSE loss instead of MAE loss.}

\begin{figure}
    \centering
    \includegraphics[width=\linewidth]{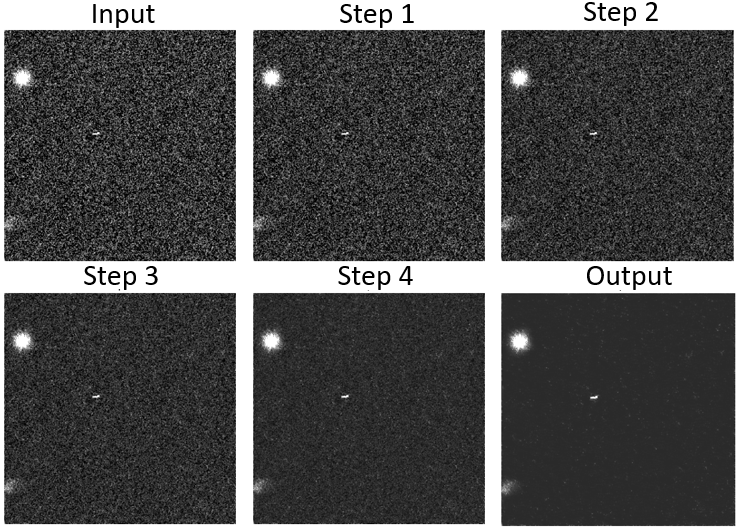}
    \caption{Example of the working of a diffusion model with five diffusion steps. The shown input is already pre-processed, while between step 4 and output there is a model prediction and post-processing, excluding denormalisation.}
    \label{fig:diffusion-example}
\end{figure}

Since the model has already learned about different noise levels during training, we proceed with the standard iterative denoising scheme, starting from the observed image $(y+x)$ to produce a clean image $(\hat{y})$. An example of this iterative denoising scheme is shown through an example image in Figure \ref{fig:diffusion-example}.

Before feeding the optical images into BGRem, they undergo pre-processing steps. Since the input images are large in size $(10560\times 10560)$, BGRem first divides them into $256 \times 256$ pixel sub-images. To handle edge effects, these cutouts overlap so that only the middle $200\times 200$ pixels are used in the final output image. This technique improves performance near the boundaries of the cutouts, significantly reducing the likelihood of sharp edges appearing at the boundaries. 

After the last diffusion step, the image undergoes two post-processing steps. First, all pixels with values below zero are set to zero. Effect of this and potential workaround to build catalog pipeline with BGRem is discussed in the Results section (Sec. \ref{sec:results}). Second, the image is denormalized using the same factor applied during the pre-processing step.

%Here, we highlight that BGRem is meant for a denoising module within a larger catalog-building pipeline, where individual source fluxes would be calculated. By clipping negative pixel values, typically resulting from noise fluctuations, one can ensure non-negative flux estimates later down the analysis pipeline. This may induce some biases for the very faint (low SNR) sources, and this is explored in Appendix (Section \ref{sec:App2}). 

\rev{During training, our model starts with the simulated ground truth image $(y)$ and iteratively adds scaled background noise $(\sigma _t x)$ as described in Eq. \ref{eq:add-noise}. As shown in Eq. \ref{eq:ddpm-forward}, the original DDPM work adds noise $(\epsilon)$ via a noise scheduling parameter $(\alpha)$, during the forward process to completely destroy the original image to produce a Gaussian Noise of zero mean and unit variance $(x_T\sim \mathcal{N}(0, \mathbb{I}))$, while our aim is to train a network under supervised scheme that can predict different levels of background noise.} 

\begin{align}\label{eq:ddpm-forward}
	x_t = \sqrt{\bar{\alpha_t}}x_0 + \sqrt{1-\bar{\alpha _t}}\epsilon; \, ~ \bar{\alpha _t}:= \prod_{s=1}^{t} \alpha_s; x_T \sim \mathcal{N}(0, \mathbb{I})  
\end{align}

\begin{figure*}
	\centering
	\includegraphics[width=0.46\textwidth]{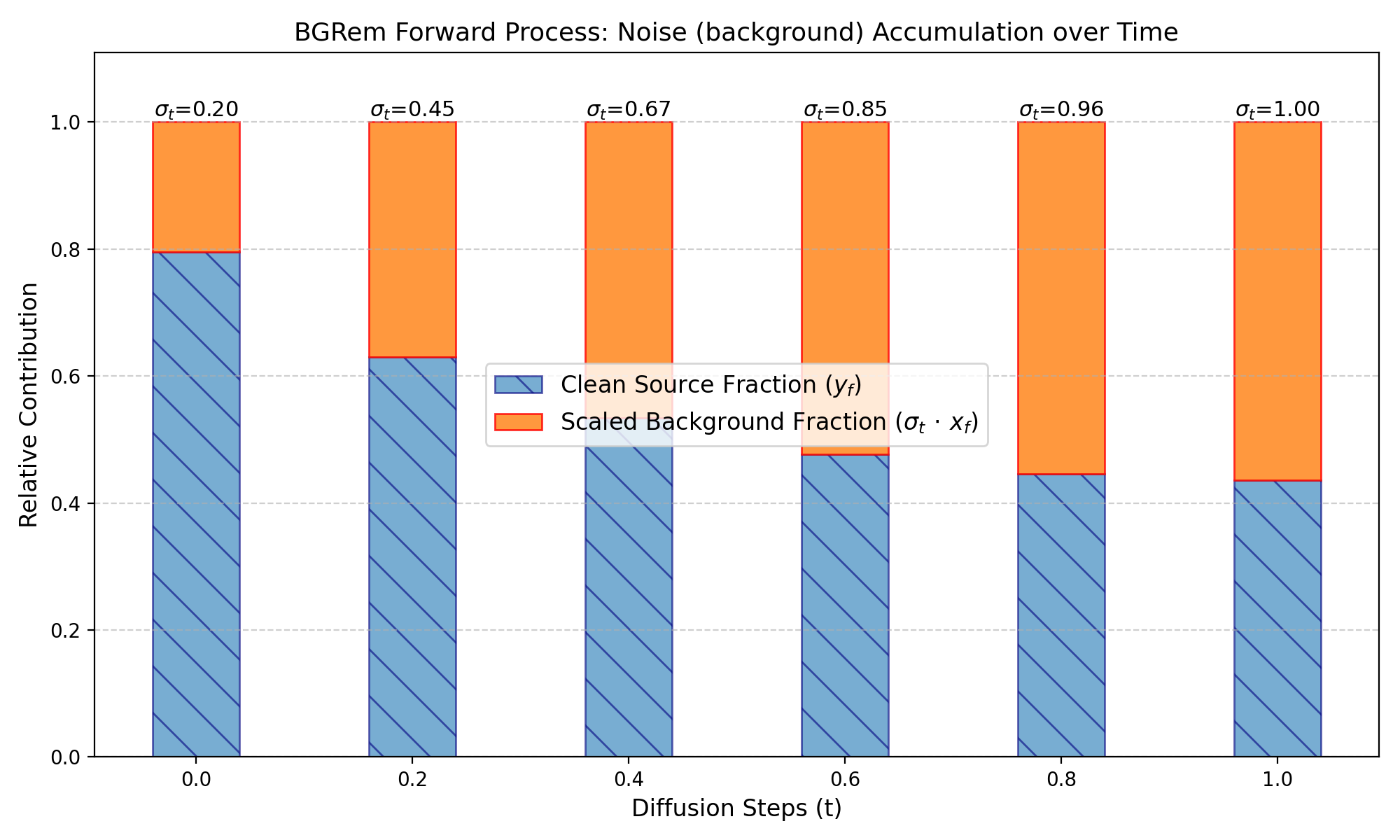}
	\includegraphics[width=0.46\textwidth]{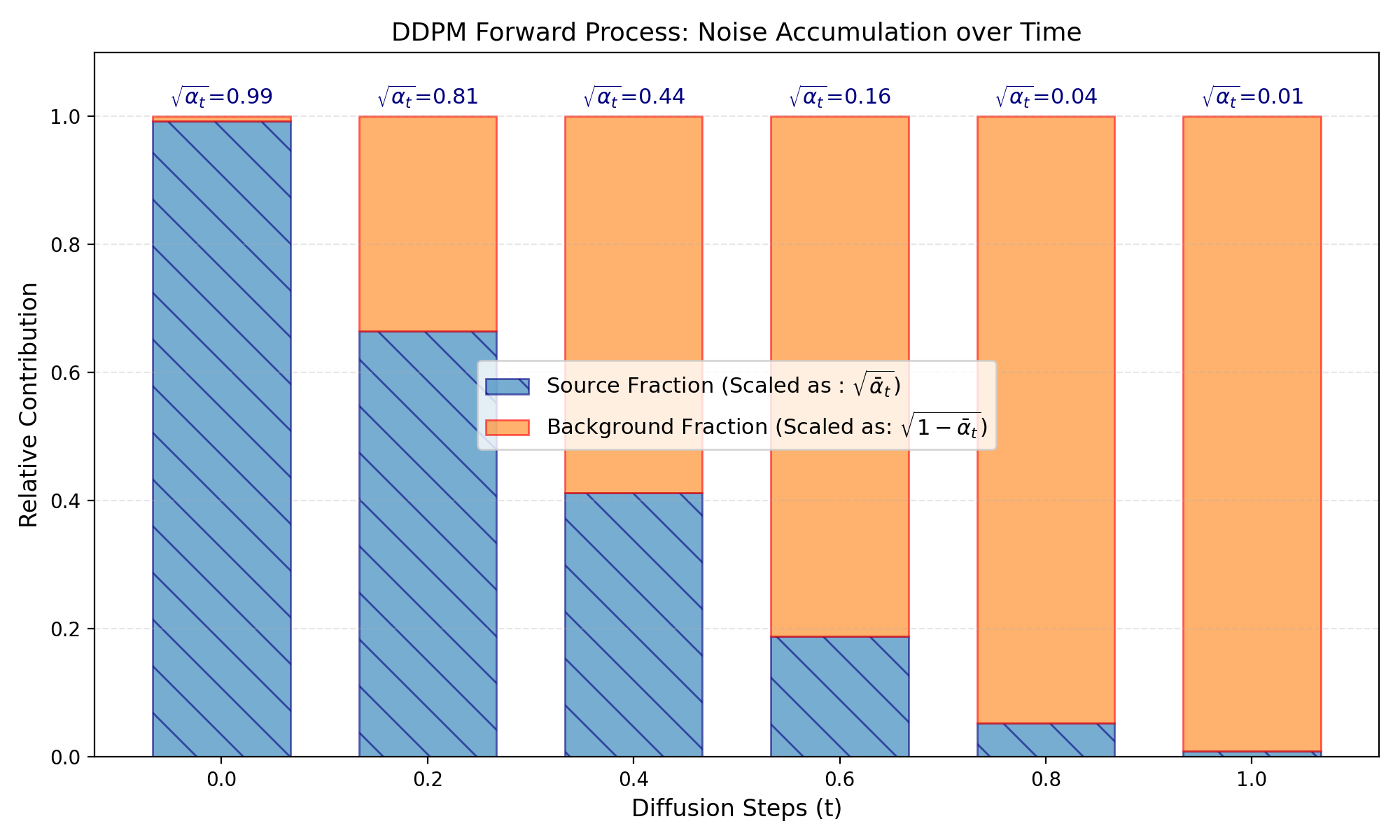}
	\caption{\rev{Comparison of forward diffusion schedules in BGRem (left, this work) and in original DDPM scheme (right, \cite{jon-ho-diffusion}). BGRem demonstrates supervised additive corruption of image with scaled background noise $\sigma_t x$ according to the noise scheduling of $\sigma _t$ described in Eq. \ref{eq:sigma-sched}. The bars represent relative contribution of photon counts from source fraction (blue hatched) and noise fraction (yellow) for a given time step, as obtained from a random image.}} 
	\label{fig:diffusion-model-difference}
\end{figure*}

\rev{To further highlight the difference, we plot schematics (Figure \ref{fig:diffusion-model-difference}) of the noise addition procedures in BGRem and in DDPM over 6 steps. For this, we take a random image pair (clean source and background only) and represent the forward process in terms of the relative photon count contributions of the signal and noise components as below: }
$$\left(\text{source (noise) counts} = \frac{\text{source (noise) counts}}{\text{source counts + noise counts}}\right)$$ 
\rev{In the standard DDPM framework, the forward process is destructive and there is no use of ``background''. The original source signal is scaled down by a factor $\sqrt{\bar{\alpha _t}}$ (Eq. \ref{eq:ddpm-forward}) and gradually replaced by stochastic Gaussian noise until the signal is effectively destroyed at $t=1.0$. In contrast, for our BGRem, we make use of the background noise and the process is additive and supervised. The clean source image $(y)$ remains constant (photon counts) throughout the schedule while a real background image $x$ is superimposed according to the sine-based scaling factor $\sigma _t$ (Eq. \ref{eq:sigma-sched}). Because the source image is never scaled down (i.e. the photon counts), the decreasing blue fraction in Figure \ref{fig:diffusion-model-difference} for BGRem, represents a relative shift in contribution caused by the accumulation of background photons, rather than physical destruction of the signal. Additionally, for our case, at the start of the noise schedule at $t=0$ with a baseline background contribution $\sigma _0 \approx 0.2$ is used to ensure the network is consistently trained against a realistic noise floor.}

When making predictions, BGRem starts with the noisy image and iteratively removes the noise. \rev{We further highlight clearly the training and inference steps for BGRem in Table \ref{tab:BGRem-train-infer}}. The choice of number of diffusion steps for inference is a hyperparameter and we discuss an optimal strategy in Appendix \ref{sec:app-diff-steps}.

%\begin{figure}
%  \centering
%  \subfloat{\includegraphics[scale=0.38]{./standard-diffusion-model.png}}
%  \subfloat{\includegraphics[scale=0.38]{./BGRem-diffusion-model.png}}
%  \caption{The fraction of noise and ground truth per diffusion step for the standard diffusion model (left) and the modified diffusion model during denoising for seven diffusion steps and noise levels equal to the image for BGRem (right). On the y-axis is the fraction of the total image flux, and on the x-axis is the diffusion step.} \label{fig:diffusion-model-difference}
%\end{figure}

% A schematic diagram for the training procedure is illustrated in Figure \ref{fig:diffusion-model-schematic}, where noise is continuously added and the U-net attempts to remove it.

\begin{table*}
	\caption{\rev{Description of the training and inference steps in BGRem}.}
	\centering
	\begin{tabular}
		{cc}
		\hline \hline
		\rev{Training} & \rev{Inference} \\ 
		\hline
		\\
		\rev{Sample a clean source image $y$} & \rev{Start from observed noisy image $x_T$}\\
		\\
		\rev{Sample a noise level $t\sim U(0, 1); \, \sigma (t) = \sin (\theta _t)$} & \rev{Set diffusion steps $t = T, \hdots , 1$} \\
		\\
		\rev{Sample background noise $x$} & \rev{At each step condition on $t$}\\
		\\
		\rev{Construct noisy image $x_t = y + \sigma_t \cdot x$} & \rev{Predict noise $\hat{x}_{\theta}$}\\
		\\
		\rev{Condition Attn. U-Net on noise level $t$} & \rev{Estimate clean image $\hat{y}$}\\
		\\
		\rev{Predict noise $\hat{x}_{\theta}(x_t, t)$} & \rev{Compute next state $x_{t-1}$}\\
		\\
		\rev{Minimize L1  loss (Eq. \ref{eq:lossl1}) } & \rev{Repeat until $t=0$} \\
		\\
		\rev{Learn denoising for all noise levels} & \rev{Output final denoised image} \\
		\\
		\hline
	\end{tabular}
	\label{tab:BGRem-train-infer}
\end{table*}

%\begin{figure}
%    \centering
%    \includegraphics[width=0.8\linewidth]{./Diffusion-model-schematic2.png}
%    \caption{Schematic of the diffusion model used in BGRem for 3 diffusion steps. The forward noise addition is only used during training. When making predictions, the model goes from right to left, iteratively removing the noise.}
%    \label{fig:diffusion-model-schematic}
%\end{figure}

\begin{figure*}
    \centering
    \includegraphics[width=0.8\linewidth]{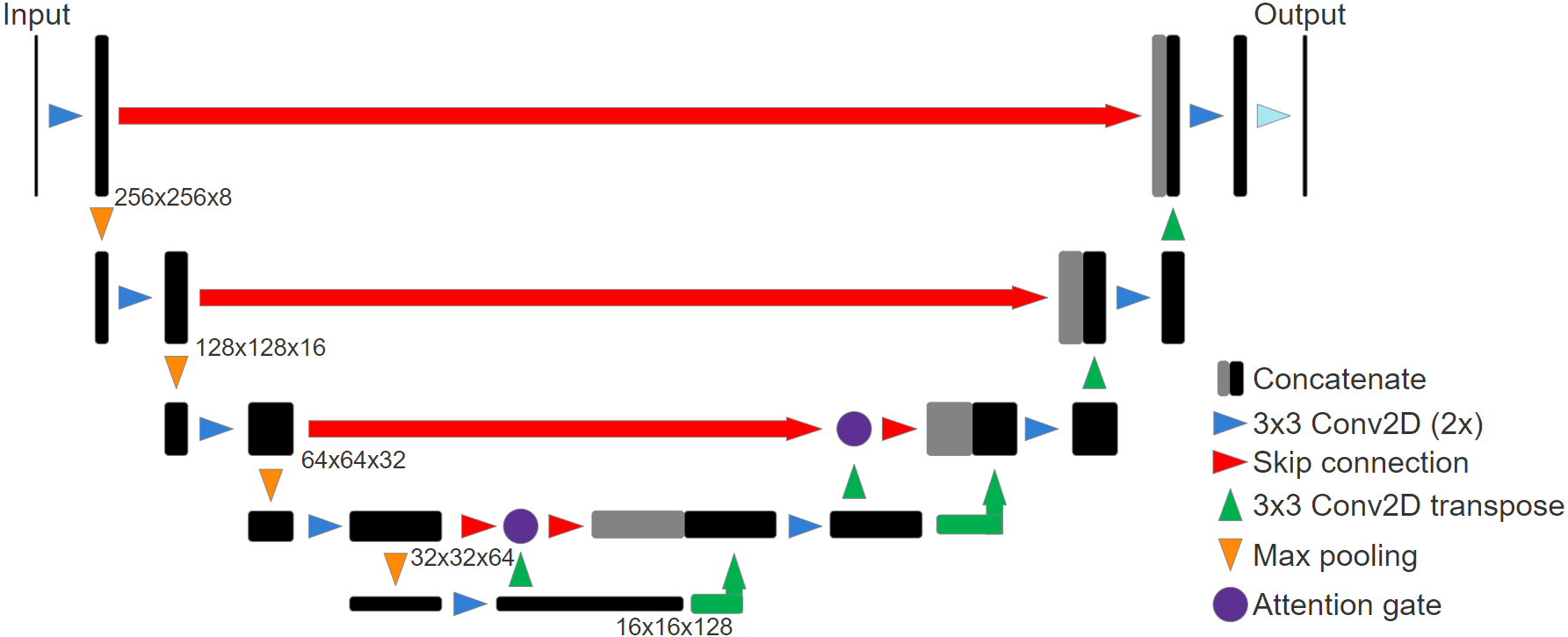}
    \caption{Schematic of the backbone architecture used in BGRem, modified from the original attention U-Net \citep{oktay2018attention}.}
    \label{fig:attention-unet}
\end{figure*}

The DNN that is used as a backbone for BGRem is an attention U-Net. This combines the successful concepts of a U-Net from \cite{u-net} and the attention mechanism from \cite{vaswani2017attention}. \cite{oktay2018attention} has already shown that the addition of attention gates improves the prediction accuracy of a U-Net with minimal computational cost. The architecture of the attention U-Net as implemented within BGRem is shown in Figure \ref{fig:attention-unet}. \rev{Here, we also highlight that in this work, BGRem has been tested under different conditions from simulated images, to real images, zero-shot testing for optical data from various telescopes and later on gamma-ray data as a part of the multi-wavelength study. Thus ``background'' in this work has been defined rather operationally, than statistically. Under the supervised scheme by providing explicit pairs of $(y, \sigma _tx)$, BGRem (i.e. the attention U-Net) learns to map the total intensity fluctuations of the background, regardless of their specific physical origin as a single component to be separated from the point-like source signal.}

The model was developed and trained in Python with the TensorFlow framework\footnote{https://www.tensorflow.org/} \citep{abadi2015tensorflow}, and its high-level Keras\footnote{https://keras.io/} API \citep{chollet2015keras}. We used LeakyRelu \citep{leaky-relu} as the activation function. The model was trained for $\sim 100$ epochs using the AdamW optimizer \citep{kingma2014adam, Loshchilov-AdamW} with a starting learning rate of $10^{-3}$ and a steadily decreasing learning rate of $0.95$, where at every epoch the learning rate gets multiplied by a factor $0.95$. Along with this, we also used early stopping to stop training if the validation loss does not improve over 10 epochs to prevent overfitting. \rev{Finally, we list all the hyperparameters for our training in the Table \ref{tab:hyperarams}.}

\begin{table*}
	\centering
	\caption{\rev{List of Hyperparameters in BGRem training and values}}
	\begin{tabular}
		{ccc}
		\hline \hline
		\rev{Hyperparameter} & \rev{Values} & \rev{Other Information}\\
		\hline
		\\
		\rev{Start Learning Rate} & \rev{$0.001 ~(0.0008)$} & \rev{MeerLICHT Data (Fermi-LAT Data)}\\
		\\
		\rev{Learning Rate Decay} & \rev{$0.95$} & \rev{--}\\
		\\
		\rev{Early Stopping} & -- & \rev{Validation Loss doesn't improve after 10 epcohs}\\
		\\
		\rev{max\_signal\_rate} & \rev{$0.0$} & \rev{$\theta _{\text{start}} = \arccos$ (max\_signal\_rate)$\approx 0.2$}\\
		\\	
		\rev{min\_signal\_rate}	& \rev{$0.98$} & \rev{$\theta _{\text{end}} = \arccos$ (min\_signal\_rate)$\approx 1.6$}\\
		\\
		\rev{Normalization Factor} & \rev{$0.01$} & \rev{Further Expanded in Appendix \ref{sec:app-diff-steps}}\\
		\\
		\rev{Diffusion Steps} & \rev{6} & \rev{Further Expanded in Appendix \ref{sec:app-diff-steps}}\\
		\\
		\rev{Optimizer} & \rev{AdamW} & \rev{Weight Decay $=1e-4$} \\
		\\	
		\rev{Batch Size} & \rev{32 (16)} & \rev{MeerLICHT Data (Fermi-LAT Data)}\\
		\\
		\rev{Start Epochs} & \rev{100 (150)} & \rev{MeerLICHT Data (Fermi-LAT Data)}\\		
		\\
		\rev{Total Parameters in Attention U-Net} & $\rev{834,801}$ & \rev{Trainable: $834,609$}\\
		\hline	
	\end{tabular}
	\label{tab:hyperarams}
\end{table*}
% \footnotetext{import tensorflow as tf}

%%%%%%%%%%%%%%%%%%%
%\input{results}
% Results
%%%%%%%%%%%%%%%%%%%

\section{Results}\label{sec:results}

To quantify BGRem's performance, we compared it to the widely used background subtraction method in SExtractor \citep{bertin1996sextractor}. SExtractor was developed to produce catalogs of astronomical sources using large-scale sky surveys, and estimation and subtraction of the background are the first steps in the pipeline. Several other recent alternatives to SExtractor, like ProFound \citep{robotham2018-profound}, NoiseChisel \citep{akhlaghi2015-noiseChisel}, were developed in recent years; however, it was shown through extensive analysis \citep{compare-tools} that SExtractor performance is comparable to these recent alternatives, and we focus on SExtractor for further analysis and comparison with BGRem. SExtractor models the background by estimating the local background by creating grids of a certain size (e.g. $64\times 64$), and computing a mode (defined as: mode $\approx 2.5 \times$ median $- 1.5 \times$ mean)\footnote{https://sourcextractorplusplus.readthedocs.io/\\en/latest/Background.html}, of the pixel values, similar to the approach used in the DAOPHOT program \citep{stetson1987daophot}. The background is then estimated via bilinear interpolation over a grid of modes in the image and subtracted from the image.

\begin{figure}
    \centering
    \includegraphics[width=\linewidth]{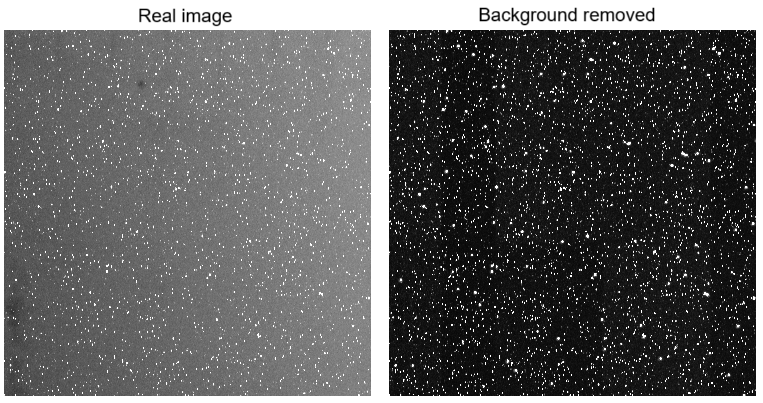}
    \caption{MeerLICHT image (left) and the image with the background noise removed with BGRem (right).}
    \label{fig: MeerLICHT BGRem} 
\end{figure}

An example of BGRem on optical images with a relatively smooth background is shown in Figure \ref{fig: MeerLICHT BGRem}. The image shows that the background gets almost completely removed, while the sources remain intact. Going beyond just visual inspection, next, we discuss our strategy to quantify BGRem's performance for the complete test set of images.

\begin{figure*}
    \centering
    \includegraphics[width=0.47\linewidth]{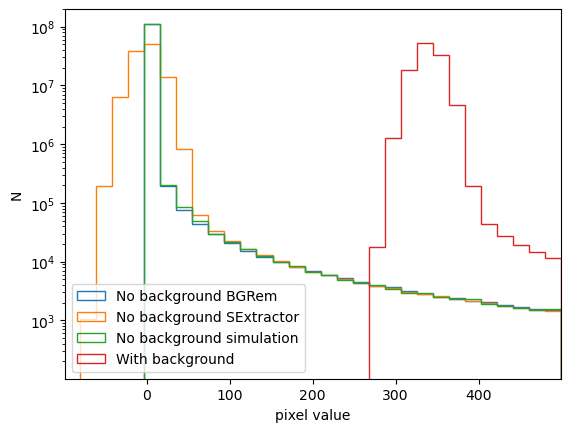}
    \includegraphics[width=0.47\linewidth]{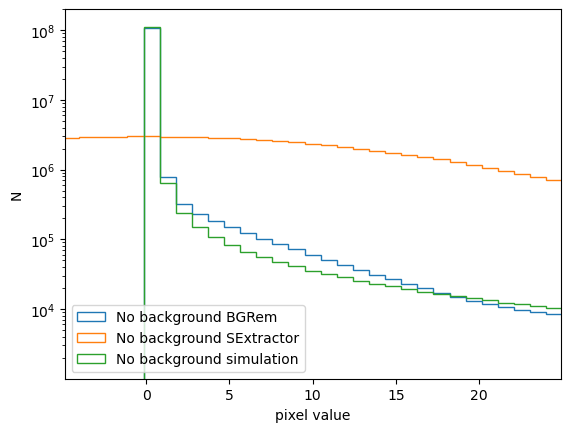}
    \caption{Left: A histogram of the pixel values for the original simulated image (red), the ground truth no background image (green), SExtractors prediction (orange) and BGRems prediction (blue). Right: Zoomed-in version of the left plot (pixel value 0-25) to highlight the close resemblance of BGRem performance with the simulation.}
    \label{fig:histogram-large-small-scale}
\end{figure*}

The distribution of the pixel values in the test images with and without the background removal by BGRem and SExtractor can highlight the effectiveness of BGRem over SExtractor, and this is shown in Figure \ref{fig:histogram-large-small-scale}. An image with stars and no background would have zero values on most pixels, with a tail of positive values, since stars only make up a small portion of the image. The left panel of Figure \ref{fig:histogram-large-small-scale} shows that for a test set of images, the histogram of pixel values obtained with BGRem closely matches this ground truth, while the histogram obtained with SExtractor is more spread out. This is shown more clearly in Figure \ref{fig:histogram-large-small-scale}, where we have zoomed in towards the low pixel values, regions of images where background removal is most effective with BGRem compared to SExtractor. 

\begin{figure*}
	\centering
	\includegraphics[width=0.47\linewidth]{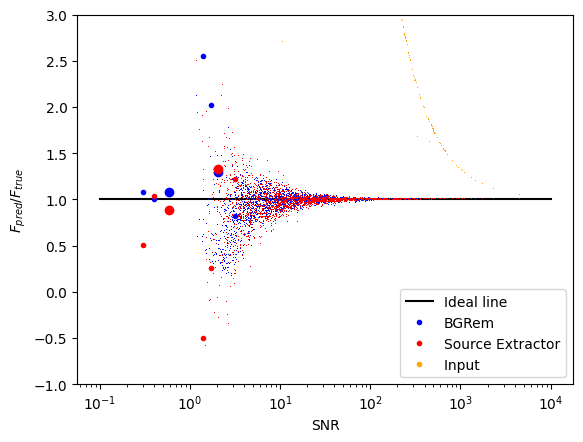}
	\includegraphics[width=0.47\linewidth]{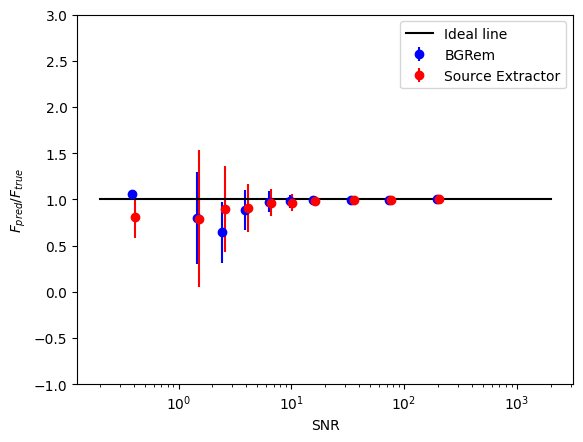}    
	\caption{Left: The predicted $\left(F_{\text{pred}}\right)$ divided by the true flux $(F_{\text{true}})$ of sources for BGRem (blue), SExtractor (red) and the input image (yellow) compared to the ideal line (black) as a function of the signal-to-noise-ratio (SNR). For every source (left), with small points being stars, dots being galaxies, and big dots being elongated galaxies. Right: Same as left, but binned in SNR, with the error bars representing statistical error.}
	\label{fig:fluxes}
\end{figure*}

\rev{As mentioned before, as a part of our post-processing step, we set all pixels with values below zero to zero. This induces a photometric bias for further downstream tasks (like source flux estimation), as with this we have shifted the mean of the background from 0 to $> 0$, which means flux estimation, especially for fainter sources, is problematic at this stage. This is highlighted in Figure \ref{fig:fluxes}, where one can clearly see the underestimation of photon counts from BGRem images compared to the photon counts from the original images. To produce Figure \ref{fig:fluxes}, first, we applied BGRem to denoise images and applied SExtractor to localize the sources. Based on the predicted locations from SExtractor, we cut a small region $(37 \times 37)$ around the predicted location as center (on the BGRem image), given our initial patch size is $256\times 256$. To determine this size, we considered two factors; first, to include larger sources (brighter and bigger), the cutout needs to be large enough; secondly, it also has to be small enough to exclude contributions from the nearby sources. We then show the sum of the counts within this patch as $F_{\text{pred}}$ as a proxy for measuring flux. $F_{\text{true}}$ represent the same, considering the source-only cutout of the same shape. This clearly highlights that at current stage BGRem cannot be used as a standalone tool to produce photometry-ready images. We intend to use BGRem for catalog-building tasks in two steps. First, the BGRem output is used solely to generate a high-precision detection mask (building upon our previous work in \cite{stoppa2022autosourceid-L}). This mask is then applied to the original, unclipped, background-subtracted data for final flux integration. This ensures that the statistical symmetry of the noise is preserved, providing better photometry while benefiting from a larger number of source detections (highlighted later in source localization performance plots). We also highlight that previously, we have developed a more dedicated DNN-based pipeline for calculating the flux of individual sources \citep{stoppa2023autosourceid-FE}; however, this simple aperture photometry approach provides a sufficiently accurate proxy for our background noise removal analysis, as the full DNN-based approach is not the target of this work.}

\begin{figure}
    \centering
    \includegraphics[width=\linewidth]{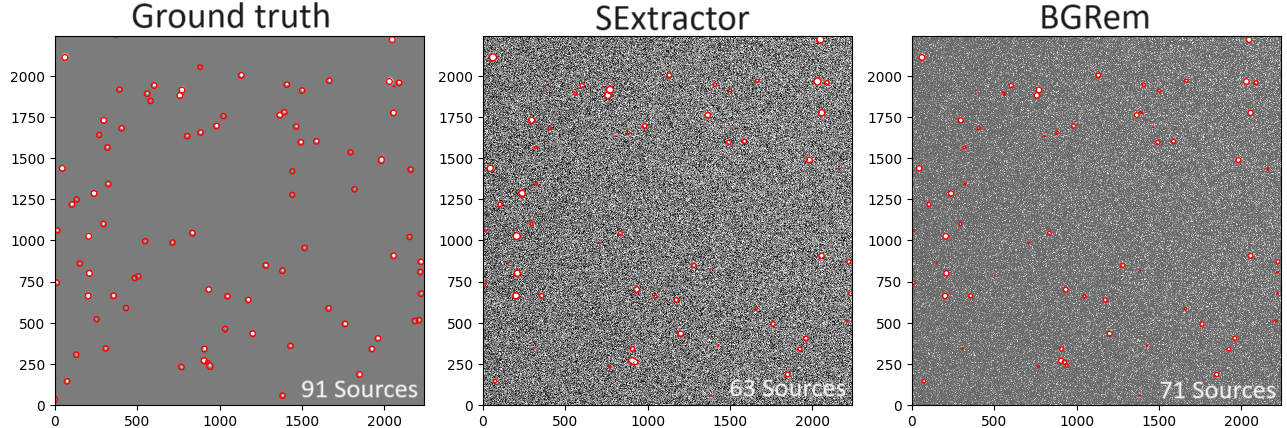}
    \caption{Sources found by SExtractor in the ground truth image (Left), the SExtractor background subtracted image (Middle), and the BGRem background subtracted image (Right). The red circles represent the predicted shape and size of the detected sources.}
    \label{fig:source-localisation}
\end{figure}

\begin{figure}
    \centering
    \includegraphics[width=0.9\linewidth]{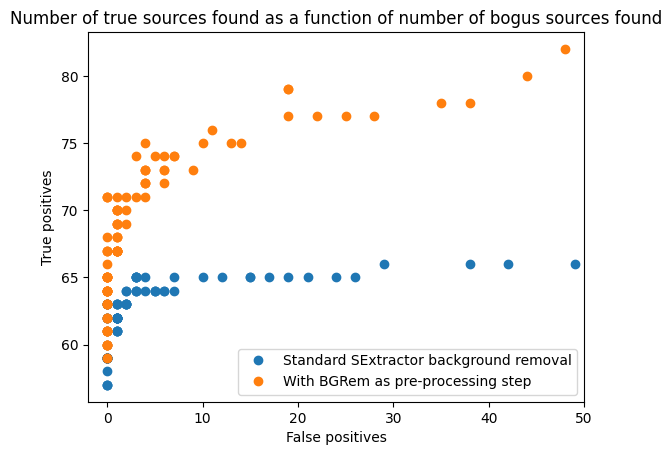}
    \caption{The number of true positives (number of sources correctly found) as a function of the number of false positives (the number of bogus sources found). The dots represent different runs of the source localization with different thresholds in source significance in SExtractor. This was done on a random cutout in the test data.}
    \label{fig:true-vs-bogus-sources}
\end{figure}

Using BGRem as a pre-processing step can improve the performance of other applications. The source localization of SExtractor can detect more sources if BGRem is used as a pre-processing step. This is shown on a cutout in the test data. Figure \ref{fig:source-localisation} shows the most sources (true positives) that could be found without any false positives, and clearly shows that by using BGRem, eight more sources could be found in this particular example. It is also informative to see how the model performs on different values of the threshold (\texttt{thresh}) parameter in SExtractor, which controls the minimum significance level to identify sources. When this threshold value is low, we expect to detect fainter sources (more true positives), but the number of false positives will also increase and vice versa. This is explored in Figure \ref{fig:true-vs-bogus-sources} and shows that using BGRem as a pre-processing step helps to find more true positives over false positives for different values of this threshold. 

\begin{figure}
  \centering
  \subfloat[]{\includegraphics[scale=0.39]{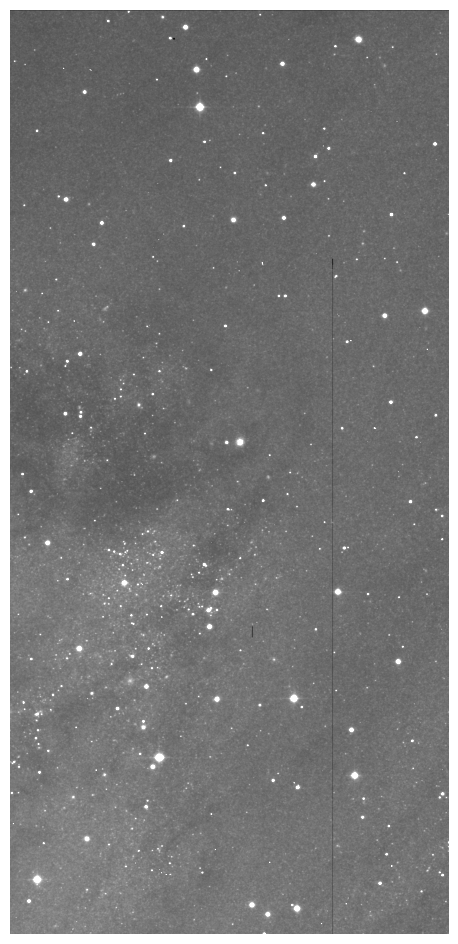}}
  \subfloat[]{\includegraphics[scale=0.39]{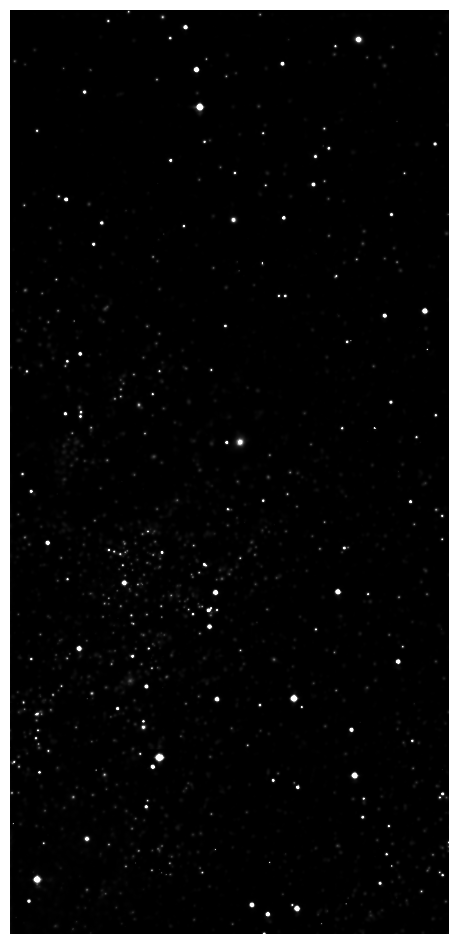}}
  \caption{An example of zero-shot inference of BGRem is shown here. The best model obtained from training and testing with MeerLICHT data is directly used for CFHT telescope image (a) to remove the background noise image with BGRem to obtain the denoised image (b). Image credit: CFHT / MegaCam (2022), operated by NRC Canada, CNRS France \& University of Hawaii. Observations from Maunakea with care and respect.} 
  \label{fig:weird-background-bgrem}
\end{figure}

BGRem also performs well on images with more complex backgrounds and smaller sizes, and here we show an example of its generalizability for a different domain of data in Figure \ref{fig:weird-background-bgrem}. This image is taken by the MegaCam of the Canada-France–Hawaii Telescope\footnote{https://www.cfht.hawaii.edu/} (CFHT) in the $g$-band \citep{CFHT-lens} and it features a part of the M31 galaxy in the background, making background removal challenging. \rev{While BGRem isolated point sources with high precision, it simultaneously demonstrated a clear limitation: the model interpreted the low-surface-brightness (LSB) structures of M31 as part of the background component and almost completely suppressed them. This aggressive removal of extended structure highlights a domain gap; because our model was trained on MeerLICHT data dominated by point sources, the attention U-Net learned to categorize any faint extended emissions as `noise'. Consequently, while the model shows impressive generalizability in removing instrumental and sky backgrounds, it is currently unsuitable for the preservation of extended galaxies or diffuse nebular structures. This underscores that BGRem, in its current iteration, is primarily intended for enhancing point-source detection tasks. Since our model is trained for MeerLICHT data and no fine-tuning is applied to adapt our model for this new data domain, the effectiveness in preserving the point sources highlights that the model learned generalizable features. The possibility of significant improvement with fine-tuning or training a separate BGRem model from scratch for this particular dataset is currently beyond the scope of this paper. Future iterations involving multi-scale loss functions or training sets with injected extended sources would be necessary to ensure morphological preservation for LSB science.}

\begin{figure*}
    \centering
    \begin{minipage}[t]{0.45\textwidth}
        \centering
        Original Image\\[-0.3ex]
        \includegraphics[width=\textwidth]{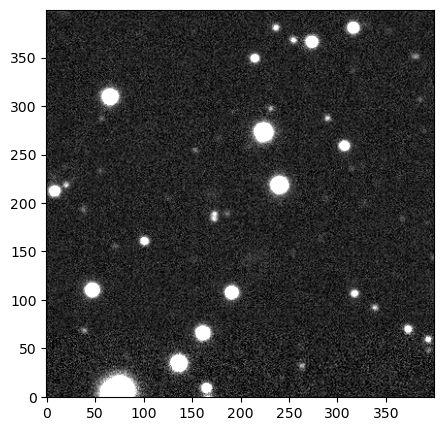}
    \end{minipage}
    \hfill
    \begin{minipage}[t]{0.45\textwidth}
        \centering
        Background Removed\\[-0.3ex]
        \includegraphics[width=\textwidth]{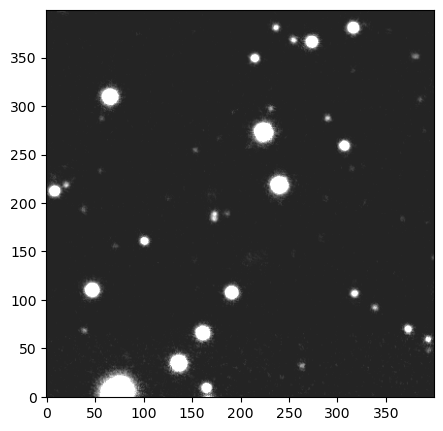}
    \end{minipage}
    \caption{Left: Part of the cutout from the Legacy Survey used for source localisation results shown in Figure~\ref{fig:true-vs-bogus-sources-real-image}; Right: Result of applying BGRem to remove background on the left image.}
    \label{fig:legacy-image}
\end{figure*}

\begin{figure}
    \centering
    \includegraphics[width=\linewidth]{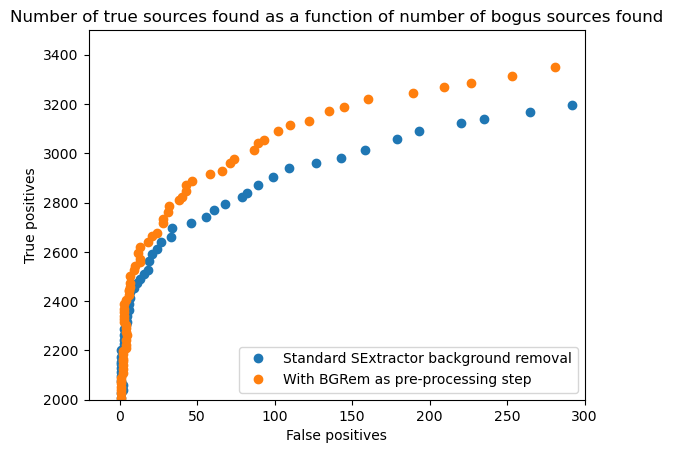}
    \caption{Same as in figure \ref{fig:true-vs-bogus-sources}, but now the application is on real data obtained from the Legacy Survey.}
    \label{fig:true-vs-bogus-sources-real-image}
\end{figure}

To quantify the generalizability of BGRem on a different dataset than it was trained on, here we used data from the Legacy Survey \citep{Dey2019-Legacy} taken with the Blanco Telescope DECam instrument, as a part of zero-shot inference testing. We used a FITS image retrieved from the Legacy Surveys DR9 Sky Viewer tool\footnote{\url{http://legacysurvey.org/viewer}} to test our ML algorithm. The image, partly shown in Figure \ref{fig:legacy-image}, is derived from the DECaLS portion of the DESI Legacy Imaging Surveys from \cite{dey2019overview}, with and without the application of BGRem. Similarly to Figure \ref{fig:true-vs-bogus-sources}, we plot the number of true positives as a function of the number of false positives in Figure \ref{fig:true-vs-bogus-sources-real-image}. Even on a large cutout of real data from the Legacy Survey, here we can quantify the zero-shot inference capability of BGRem and highlight that throughout the different levels of threshold values in SExtractor, BGRem performs overall better than the SExtractor background removal tool. Here, we also highlight that for both Figure \ref{fig:true-vs-bogus-sources} and Figure \ref{fig:true-vs-bogus-sources-real-image}, we tested the results by not clipping the pixel values to zero when negative values are predicted with BGRem. The performance of SourceExtractor to detect sources gets worse in both cases.

Finally, here we highlight the possibility of training BGRem to denoise astronomical images at a different wavelength with a different noise component than optical images, as previously described. We focus on simulated $\gamma$-ray sky images for the Fermi-LAT telescope.  

\begin{figure*}%
    \centering
    \subfloat{{\includegraphics[width=0.3\linewidth]{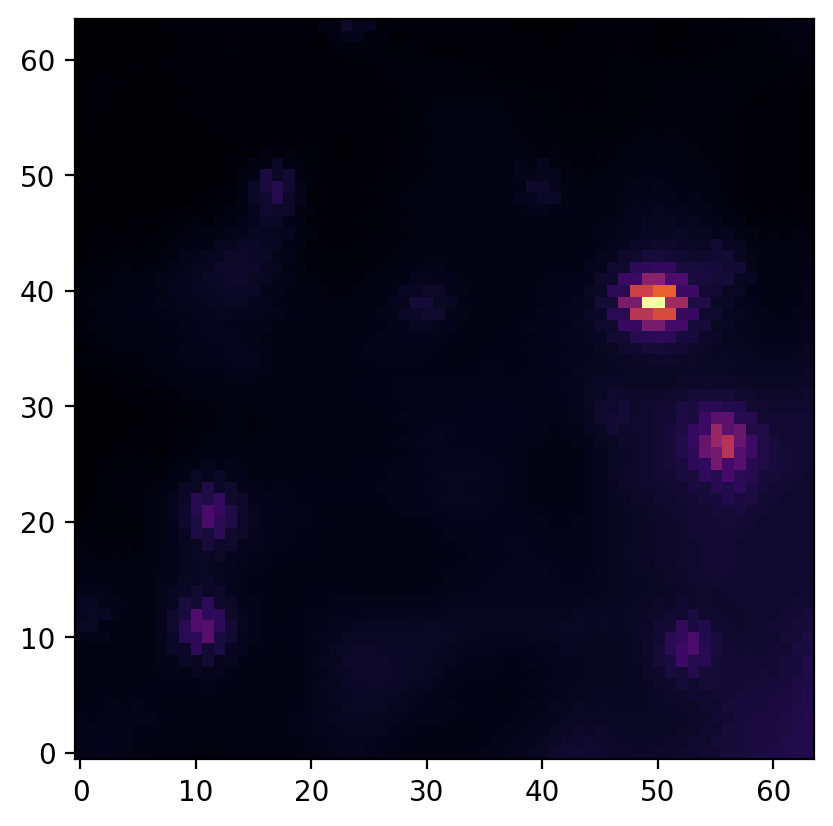} }}%
    \qquad
    \subfloat{{\includegraphics[width=0.3\linewidth]{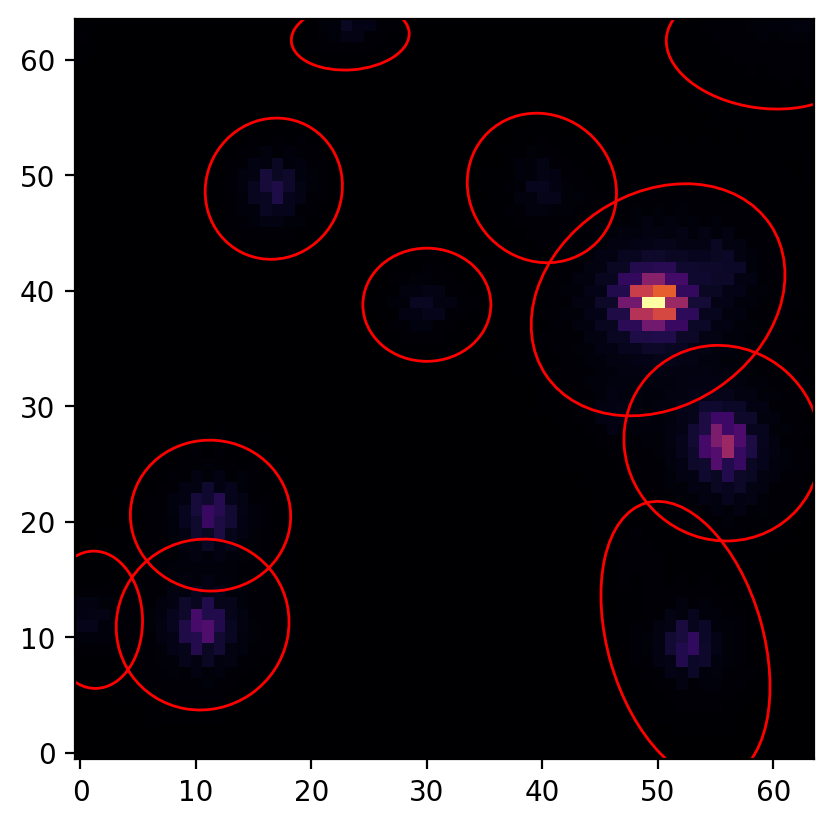} }}%
    \qquad
    \subfloat{{\includegraphics[width=0.3\linewidth]{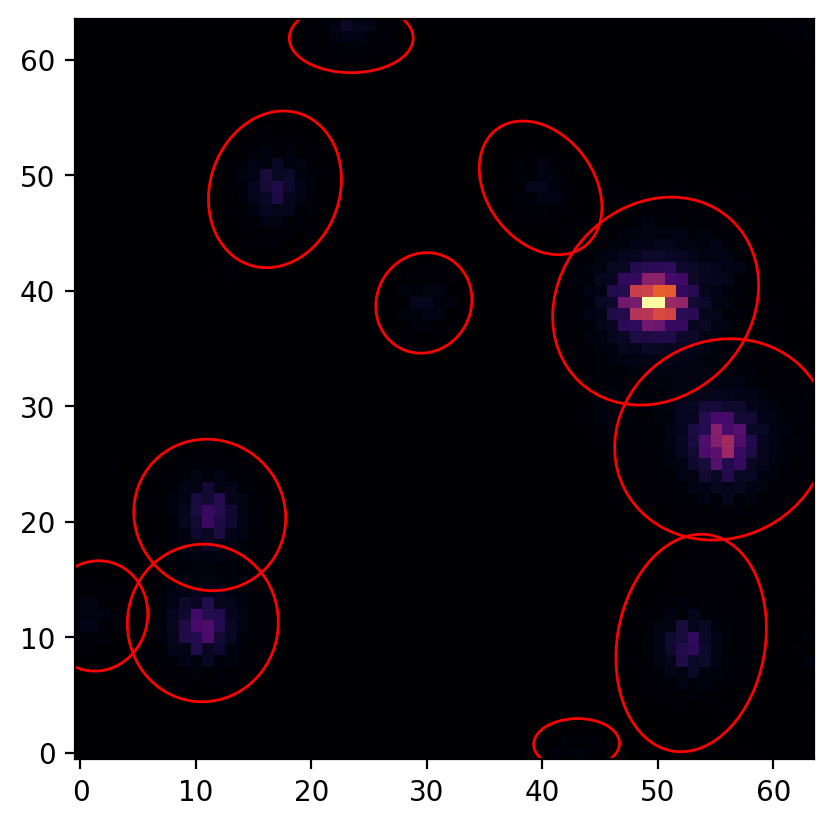} }}%

    \setcounter{subfigure}{0} % Reset subfigure counter
    
    \subfloat[\centering with Background]{{\includegraphics[width=0.3\linewidth]{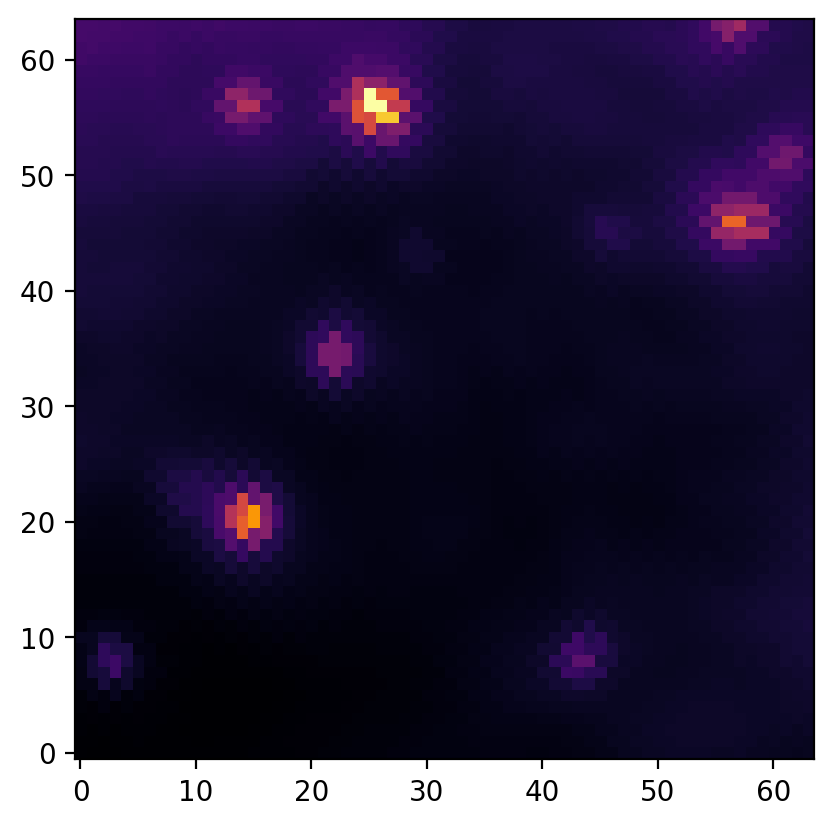} }}%
    \qquad
    \subfloat[\centering Source Only]{{\includegraphics[width=0.3\linewidth]{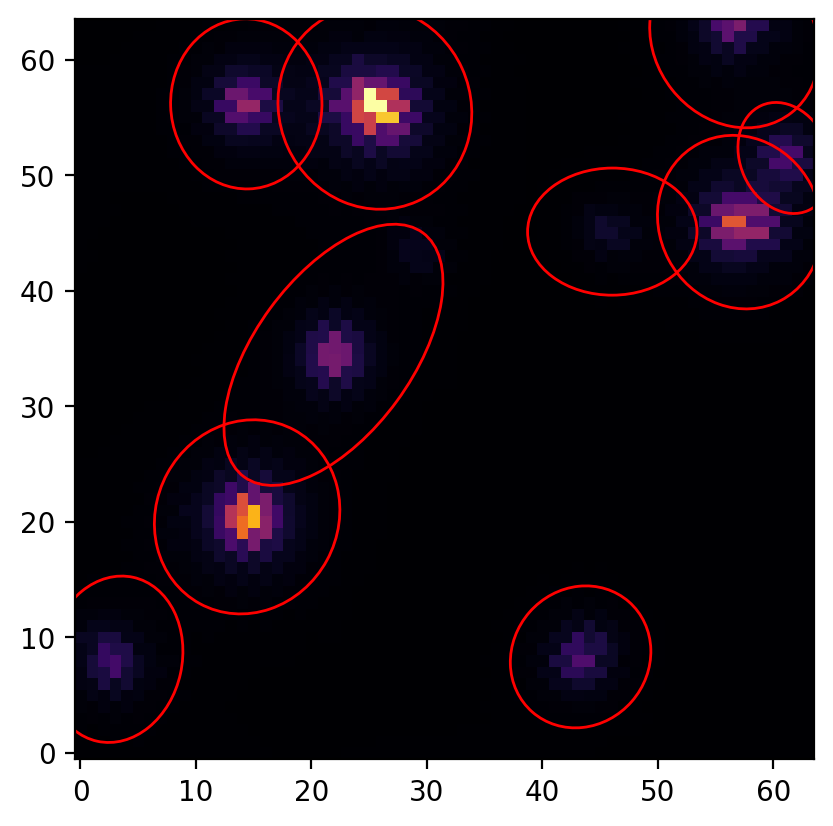} }}%
    \qquad
    \subfloat[\centering with BGRem]{{\includegraphics[width=0.3\linewidth]{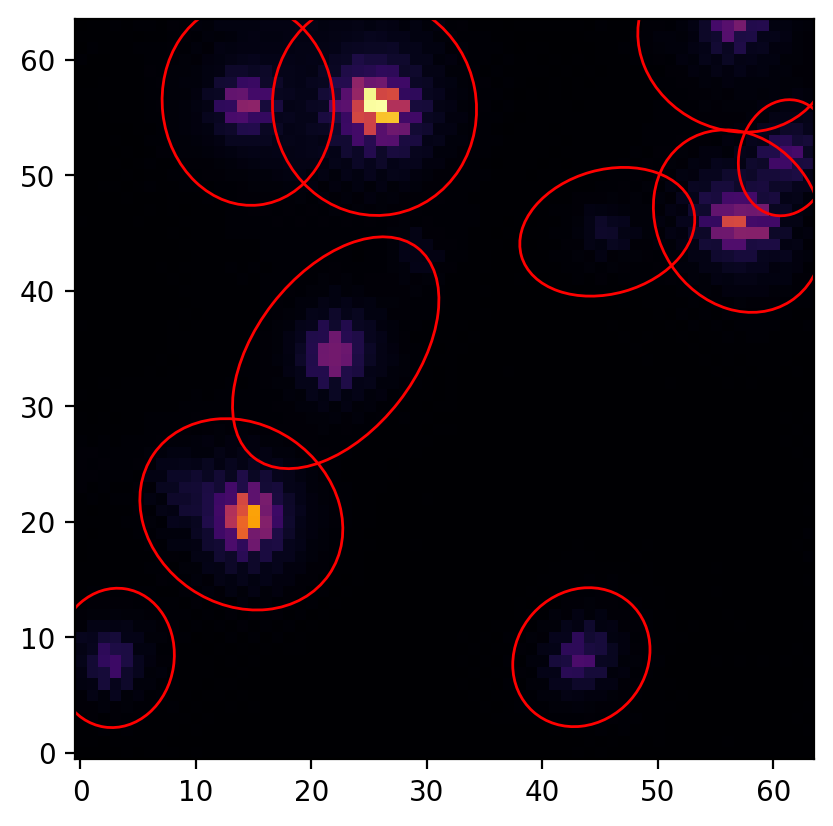} }}%
    \caption{Examples of BGRem for simulated gamma-ray sky patches are shown here for two different regions of the sky. On the top panel, the example patch is away from the Galactic plane where IEM is low, compared to the bottom panel where the IEM component is stronger. The effectiveness of BGRem is highlighted by the equivalent performance of SExtractor on Source Only (b) and BGRem images (c).}%
    \label{fig:example-bgrem-gamma}%
\end{figure*}

\begin{figure}
    \centering
    \includegraphics[width=0.9\columnwidth]{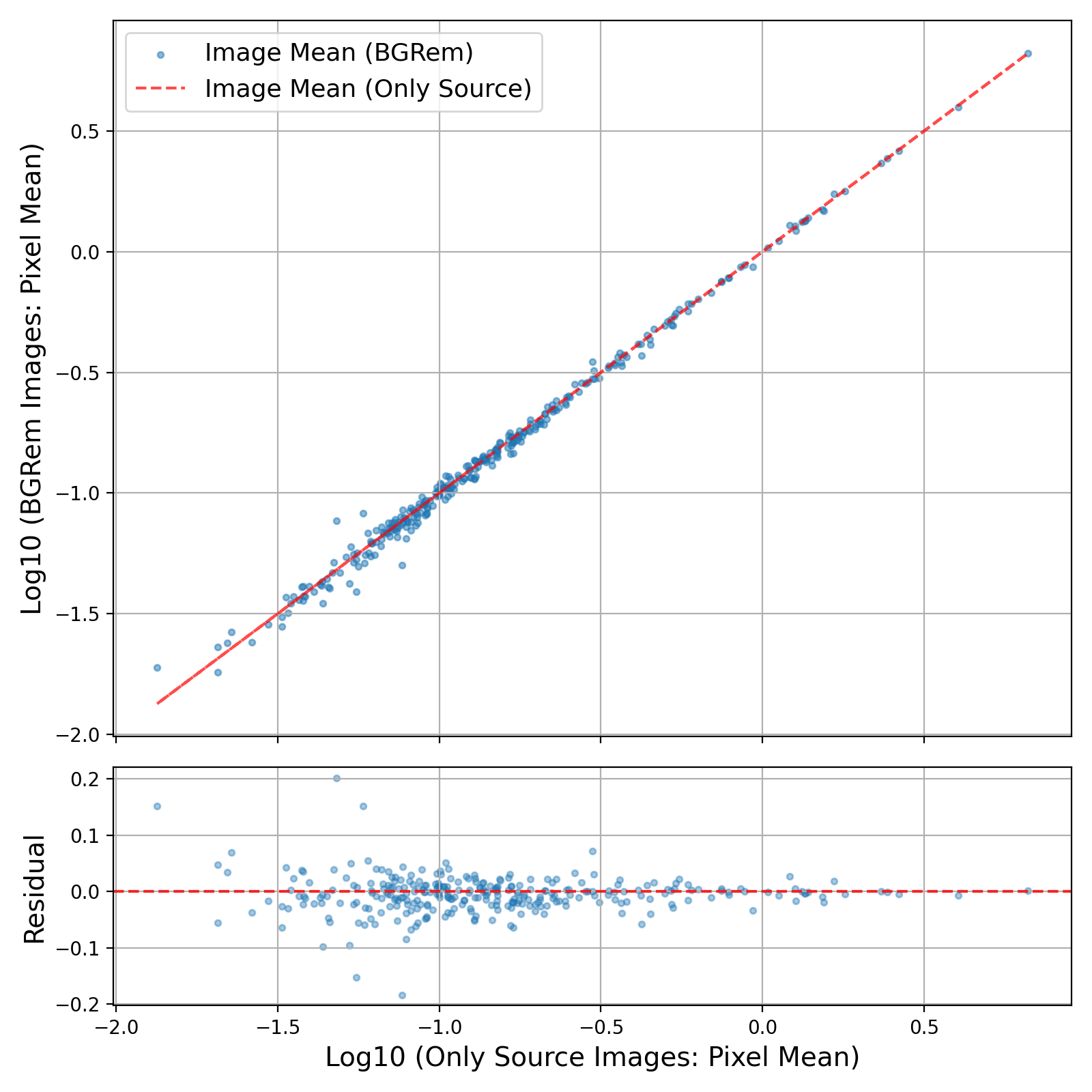}
    \caption{We show an example of mean pixel values of the denoised image patches (with BGRem) against the Source Only image patches with blue dots, which lie close to the diagonal line (red). The residual plot at the bottom panel highlights when, on average, the image patches are brighter (more and/or brighter sources), BGRem reconstruction works better.}
    \label{fig:bgrem-gamma-pix-mean}
\end{figure}

An example of BGRem performance from the test set is shown in Figure \ref{fig:example-bgrem-gamma}. Focusing on two randomly selected test patches of the sky away (close) from the Galactic Plane is shown in the top (bottom) panel of Figure \ref{fig:example-bgrem-gamma}. This highlights that the source detection rate using SExtractor on the ground truth image (source-only image) closely resembles the denoised image obtained with BGRem. We also show that in Figure \ref{fig:bgrem-gamma-pix-mean}, the mean pixel values of the denoised images with BGRem follow a diagonal line when plotted against the mean pixel values of the source-only images, highlighting the effectiveness of BGRem to reduce the contribution from the IEM noise. A low (high) mean pixel value in the source-only image would suggest the possibility of the test patch being dominated by faint (bright) sources. This highlights the reliability of reconstructing the source pixel counts with BGRem for the $\gamma$-ray images.

\begin{figure}
    \centering
    \includegraphics[width=0.90\columnwidth]{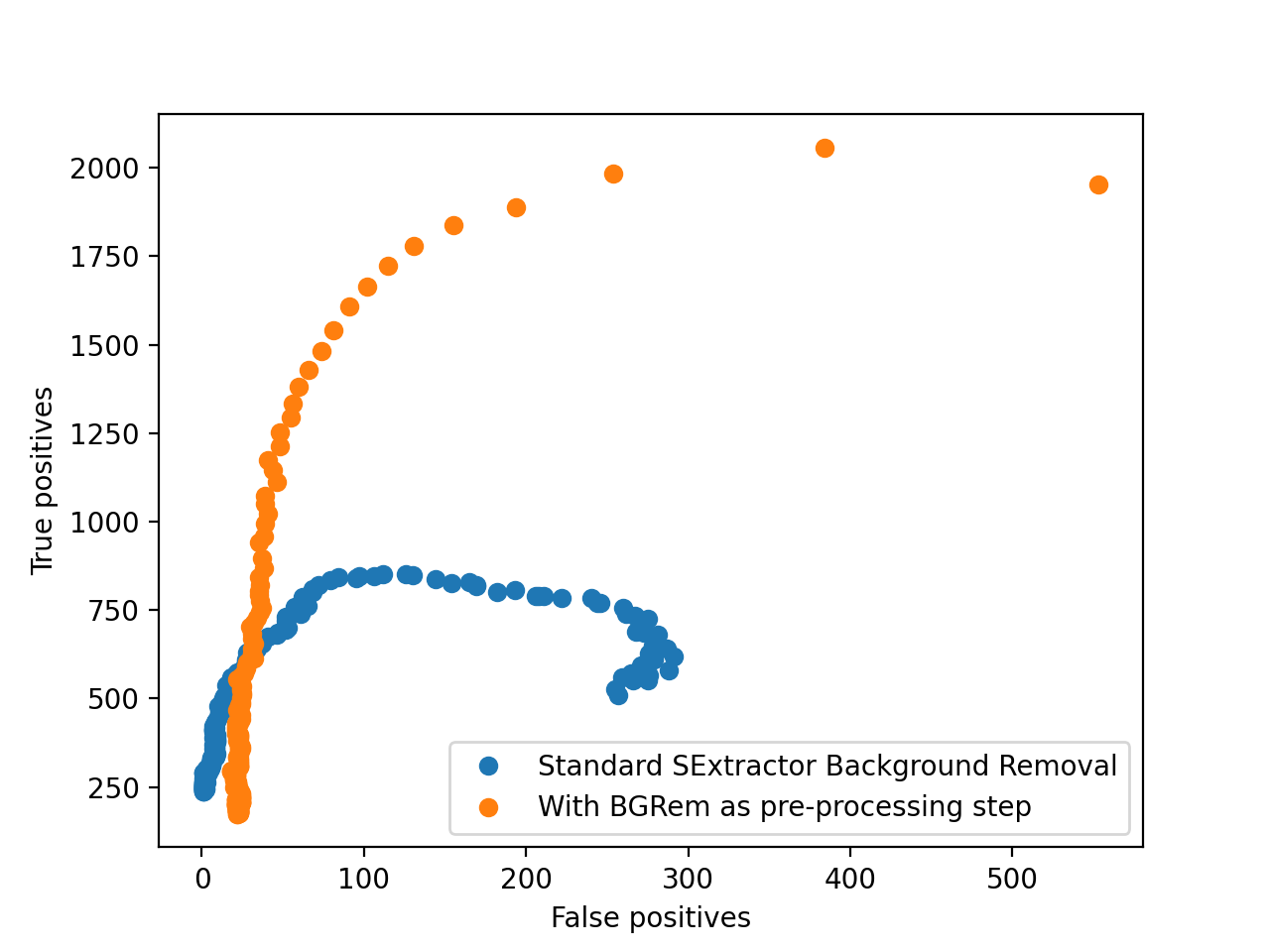}
    \caption{Same as Figures \ref{fig:true-vs-bogus-sources}, \ref{fig:true-vs-bogus-sources-real-image}, but here we show the results obtained with simulated $\gamma$-ray sky images.}
    \label{fig:TPFP-gamma-ray}
\end{figure}

As is the case with optical images, this retrained model also improves the ability of SExtractor to find sources in the images. This is shown in Figure \ref{fig:TPFP-gamma-ray}. This model seems to generate some artefacts that lead SExtractor to find false positives at very high thresholds. However, at least with the simulated data and using SExtractor as a detection module, more than double the number of correct sources can be identified by using BGRem as a pre-processing step, compared to the default SExtractor background removal method. Here, we highlight that even though SExtractor is not developed/trained for removing the IEM in $\gamma$-ray wavelength and detecting $\gamma$-ray sources, the significantly improved performance with BGRem is shown here to motivate having BGRem as a pre-processing step for current and upcoming source detection pipelines at various wavelengths. 

\begin{figure*}%
    \centering
    \subfloat{{\includegraphics[width=0.3\linewidth]{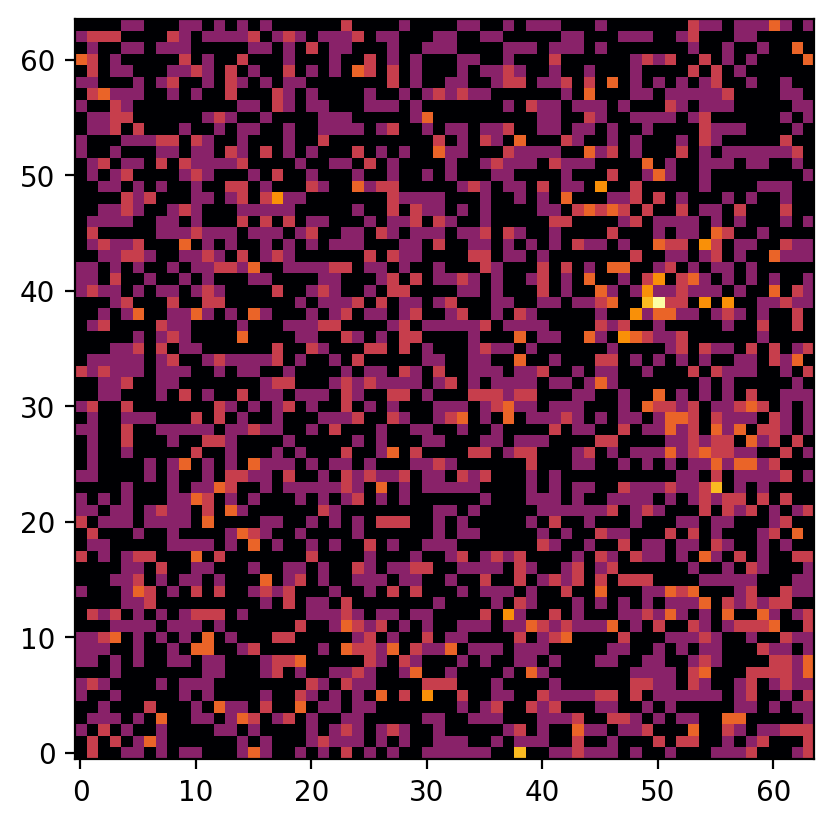} }}%
    \qquad
    \subfloat{{\includegraphics[width=0.3\linewidth]{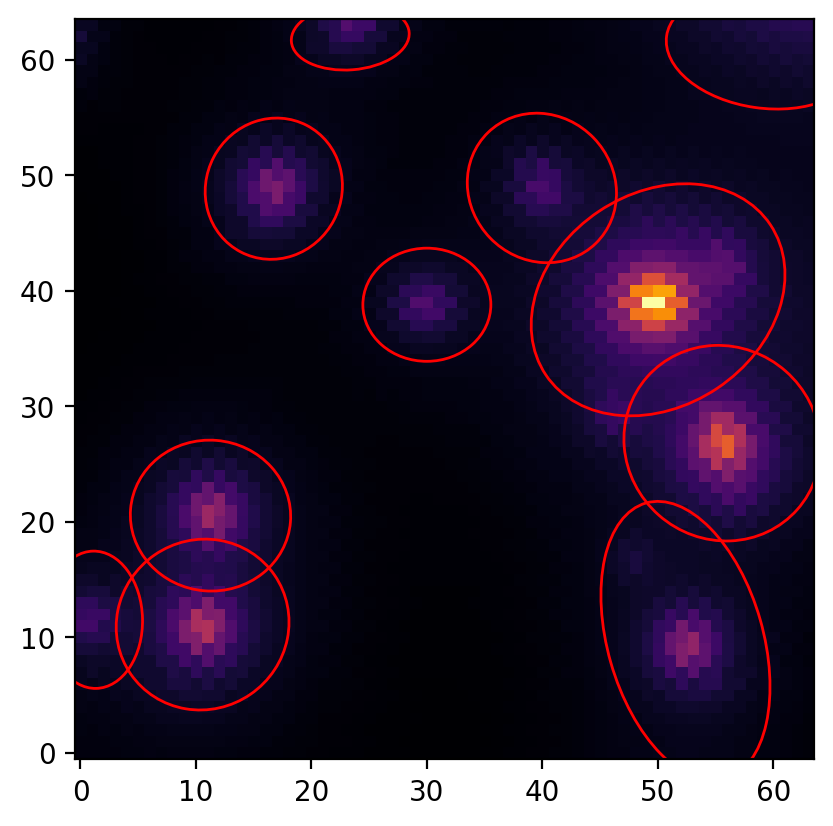} }}%
    \qquad
    \subfloat{{\includegraphics[width=0.3\linewidth]{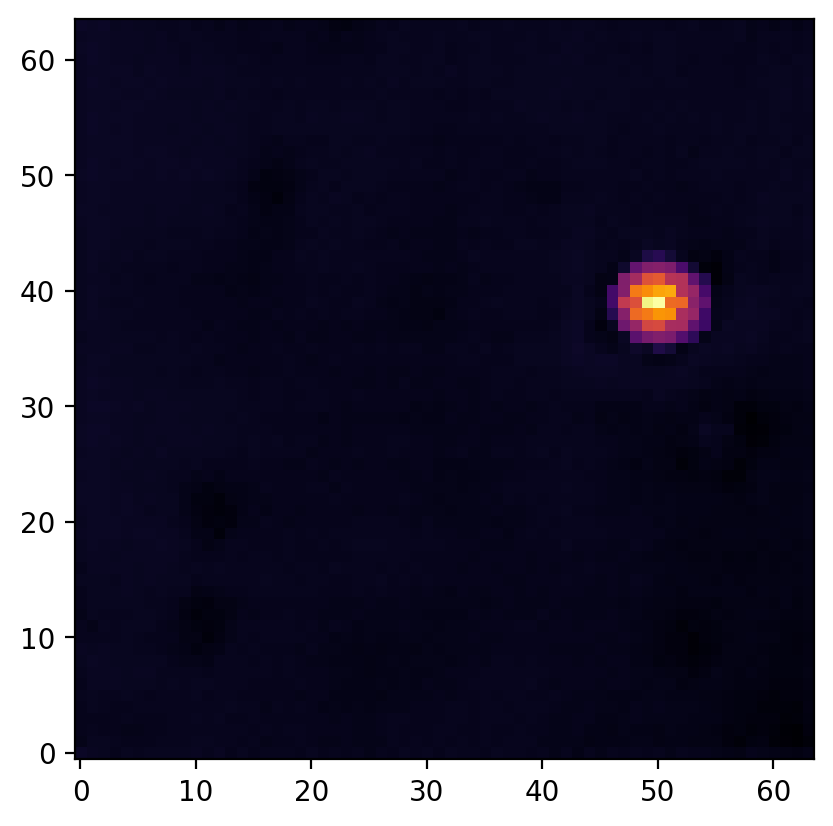} }}%

    \setcounter{subfigure}{0} % Reset subfigure counter
    
    \subfloat[\centering with Background]{{\includegraphics[width=0.3\linewidth]{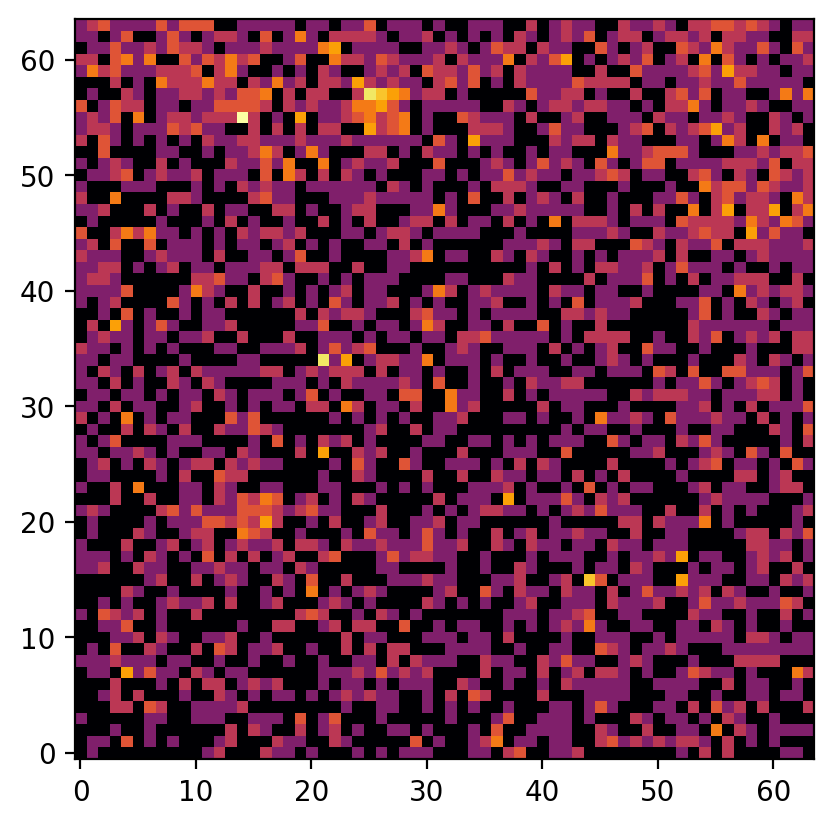} }}%
    \qquad
    \subfloat[\centering Source Only]{{\includegraphics[width=0.3\linewidth]{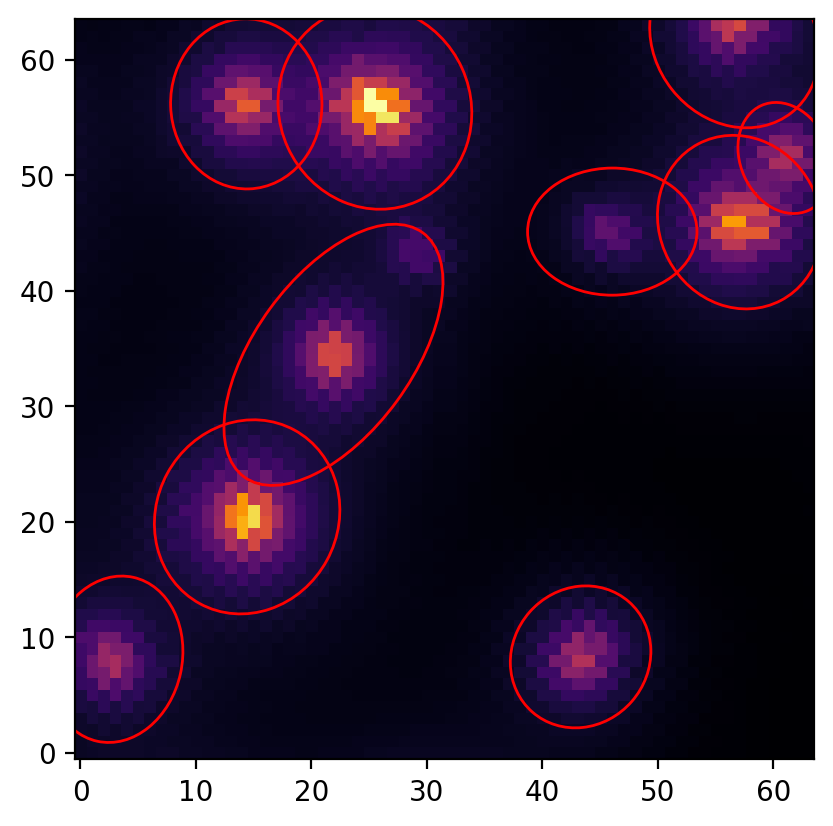} }}%
    \qquad
    \subfloat[\centering with BGRem]{{\includegraphics[width=0.3\linewidth]{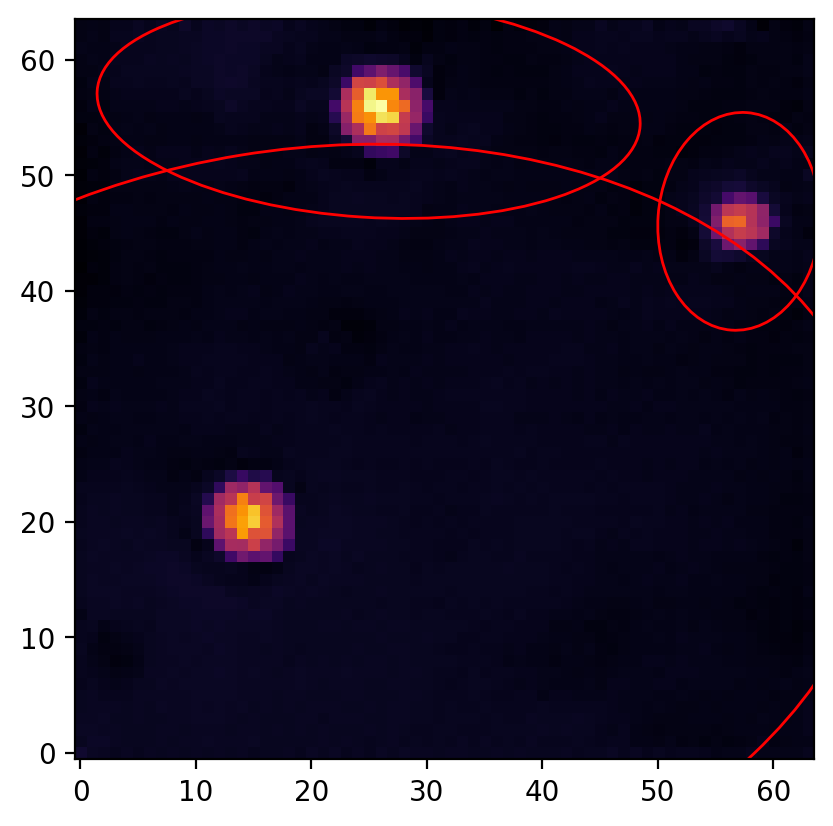} }}%
    \caption{\rev{Examples of BGRem applied for the same simulated gamma-ray sky patches as in Figure \ref{fig:example-bgrem-gamma}, but now including the poisson noise. As seen for both high and low galactic latitude patches, BGRem fails to properly remove the poisson + IEM noise, highlighting one of the drawbacks at the current stage of BGRem for $\gamma$-ray data}.}%
    \label{fig:example-bgrem-gamma-wPoi}%
\end{figure*}

\begin{figure}
	\centering
	\includegraphics[width=0.9\columnwidth]{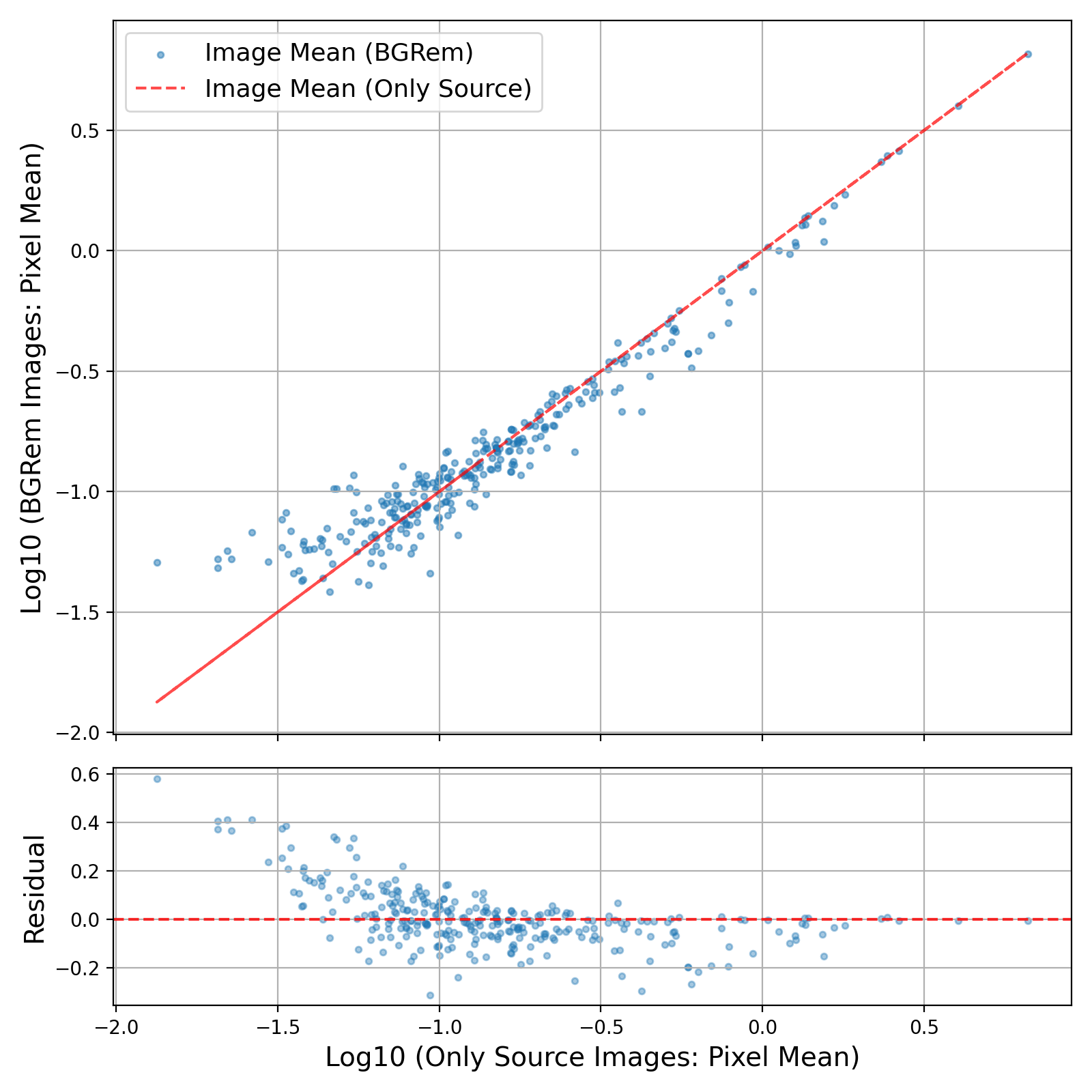}
	\caption{\rev{Same as in Figure \ref{fig:bgrem-gamma-pix-mean}, but now we highlight the effect of additional poisson noise. Clearly, BGRem overall under-predicts the counts for brighter patches and over-predicts for fainter patches}.}
	\label{fig:bgrem-gamma-pix-mean-wPoi}
\end{figure}

\rev{However, here we also highlight a current drawback of BGRem in effectively removing background noise when we include observational noise i.e. ``Poisson Noise'' for the $\gamma$-ray data. An example of BGRem applied to the same patches as shown in Figure \ref{fig:example-bgrem-gamma} but now including the poisson noise along with IEM noise is shown in Figure \ref{fig:example-bgrem-gamma-wPoi}. This clearly shows that except the brightest sources in those patches, BGRem fails to reconstruct the denoised images. To further inspect a global trend, we reproduce the Figure \ref{fig:bgrem-gamma-pix-mean-wPoi}, similar to the Figure \ref{fig:bgrem-gamma-pix-mean}, but including the effect of poisson noise for the same test-set used before, and highlight how the mean pixel values of BGRem images vary with respect to the original source-only images. Now, one can clearly see that overall BGRem performance is worse irrespective of the brightness of the images. A general trend that can be infer from this is that BGRem under-predicts the counts for brighter images, while over-predicts the counts for fainter images. For the fainter images, BGRem struggles to distinguish between a faint real source and an upward poisson fluctuation of the background. Thus, BGRem may interpret the noise spikes as signal, it won't subtract enough noise leaving behind more counts. For brighter patches, BGRem's noise prediction is more aggressive leading to subtraction of more photon counts than ideal.} 

\rev{This highlights the fact that while BGRem shows excellent promise for background removal tasks as a proof-of-concept, the observed over and under-prediction of photon counts in the poisson regime requires further investigation. Future work will focus on a systematic hyperparameter optimization, including a more robust treatment of the image normalization factors which are critical for the non-linear scaling of gamma-ray counts to ensure the model is optimized for the stochastic nature of poisson noise in addition to the structural complexities of the IEM.}

\section{Conclusion}\label{sec:conclusion}

In this paper, we introduced BGRem, a background noise remover for astronomical images. This diffusion-based model with attention U-Net as a backbone can remove the background noise from astronomical images effectively with negligible artefacts. First, we train and test BGRem for optical images from the MeerLICHT telescope, and then apply domain transfer to show the effectiveness of BGRem for optical images taken by CFHT's MegaCam and the DESI Legacy Survey on the DECam instrument. We also showed that by retraining the model on a different dataset, particularly for $\gamma$-ray simulated data for the Fermi-LAT telescope, one can have a reliable background removal tool for astronomical images across various wavelengths. The computation time also doesn't depend on the contents of the image, like the number of stars/galaxies in an image. According to our knowledge, for the first time, a diffusion-based denoising model was shown to be effective in removing background from optical and $\gamma$-ray images and help SExtractor to detect more sources. Our results demonstrate that this approach can serve as an effective pre-processing step for existing methods for building a pipeline to create an astronomical catalog. We also highlight the potential application of BGRem at a different wavelength, in $\gamma$-rays, with different background contaminations than optical images. 

\rev{However, at its current stage, BGRem possesses certain limitations that define its applicable scientific scope. As our model is trained primarily on simulations of point-like sources, it exhibits a morphological bias: the Attention U-Net backbone tends to interpret diffuse, extended galaxies as part of the background component. Consequently, BGRem may attenuate scientific signals from extended sources, making it currently unsuitable for tasks requiring surface photometry or morphological analysis of such objects. Also, BGRem tends to underestimate the flux of faint objects, as the model may misidentify low-level signals as stochastic background and remove them. Thus the direct output of BGRem should not be used for standalone high-precision photometry. While BGRem is shown to be suitable to remove IEM noise for $\gamma$-ray data, at the current stage, it fails to model and remove poisson noise.}

Future improvements could involve fine-tuning model hyperparameters, exploring alternative noise schedules or embedding strategies, and optimizing architectural components such as attention modules. Additionally, improving the quality and diversity of the training data — through more realistic background simulations or domain-specific augmentation — could help BGRem's generalizing capability even more. We aim to explore these topics in our future research. Also we aim to apply the same technique with different training data to develop a model that includes astronomical images from different wavelengths, including radio, X-ray or infrared images. While BGRem performs quite robustly across different noise levels, there remains room for further improvement.

\section*{Acknowledgements}

\rev{We would like to thank the anonymous referee for constructive comments and queries which have made the draft overall better.}

The authors are grateful to P.M. Vreeswijk for helpful comments to improve the draft. 

The work of S.B was supported by the Slovenian Research Agency under grants P1-0031, N1-0344, and J1-60014. 

PJG is partly supported by NRF SARChI grant 111692.

The work of R. RdA was supported by PID2020-113644GB-I00 from the Spanish Ministerio de Ciencia e Innovación and by the PROMETEO/2022/69 from the Spanish GVA.

The Legacy Surveys consist of three individual and complementary projects: the Dark Energy Camera Legacy Survey (DECaLS; Proposal ID \#2014B-0404; PIs: David Schlegel and Arjun Dey), the Beijing-Arizona Sky Survey (BASS; NOAO Prop. ID \#2015A-0801; PIs: Zhou Xu and Xiaohui Fan), and the Mayall z-band Legacy Survey (MzLS; Prop. ID \#2016A-0453; PI: Arjun Dey). DECaLS, BASS and MzLS together include data obtained, respectively, at the Blanco telescope, Cerro Tololo Inter-American Observatory, NSF’s NOIRLab; the Bok telescope, Steward Observatory, University of Arizona; and the Mayall telescope, Kitt Peak National Observatory, NOIRLab. Pipeline processing and analyses of the data were supported by NOIRLab and the Lawrence Berkeley National Laboratory (LBNL). The Legacy Surveys project is honored to be permitted to conduct astronomical research on Iolkam Du’ag (Kitt Peak), a mountain with particular significance to the Tohono O’odham Nation.

NOIRLab is operated by the Association of Universities for Research in Astronomy (AURA) under a cooperative agreement with the National Science Foundation. LBNL is managed by the Regents of the University of California under contract to the U.S. Department of Energy.

This project used data obtained with the Dark Energy Camera (DECam), which was constructed by the Dark Energy Survey (DES) collaboration. Funding for the DES Projects has been provided by the U.S. Department of Energy, the U.S. National Science Foundation, the Ministry of Science and Education of Spain, the Science and Technology Facilities Council of the United Kingdom, the Higher Education Funding Council for England, the National Center for Supercomputing Applications at the University of Illinois at Urbana-Champaign, the Kavli Institute of Cosmological Physics at the University of Chicago, Center for Cosmology and Astro-Particle Physics at the Ohio State University, the Mitchell Institute for Fundamental Physics and Astronomy at Texas A\&M University, Financiadora de Estudos e Projetos, Fundacao Carlos Chagas Filho de Amparo, Financiadora de Estudos e Projetos, Fundacao Carlos Chagas Filho de Amparo a Pesquisa do Estado do Rio de Janeiro, Conselho Nacional de Desenvolvimento Cientifico e Tecnologico and the Ministerio da Ciencia, Tecnologia e Inovacao, the Deutsche Forschungsgemeinschaft and the Collaborating Institutions in the Dark Energy Survey. The Collaborating Institutions are Argonne National Laboratory, the University of California at Santa Cruz, the University of Cambridge, Centro de Investigaciones Energeticas, Medioambientales y Tecnologicas-Madrid, the University of Chicago, University College London, the DES-Brazil Consortium, the University of Edinburgh, the Eidgenossische Technische Hochschule (ETH) Zurich, Fermi National Accelerator Laboratory, the University of Illinois at Urbana-Champaign, the Institut de Ciencies de l’Espai (IEEC/CSIC), the Institut de Fisica d’Altes Energies, Lawrence Berkeley National Laboratory, the Ludwig Maximilians Universitat Munchen and the associated Excellence Cluster Universe, the University of Michigan, NSF’s NOIRLab, the University of Nottingham, the Ohio State University, the University of Pennsylvania, the University of Portsmouth, SLAC National Accelerator Laboratory, Stanford University, the University of Sussex, and Texas A\&M University.

BASS is a key project of the Telescope Access Program (TAP), which has been funded by the National Astronomical Observatories of China, the Chinese Academy of Sciences (the Strategic Priority Research Program “The Emergence of Cosmological Structures” Grant \# XDB09000000), and the Special Fund for Astronomy from the Ministry of Finance. The BASS is also supported by the External Cooperation Program of Chinese Academy of Sciences (Grant \# 114A11KYSB20160057), and Chinese National Natural Science Foundation (Grant \# 12120101003, \# 11433005).

The Legacy Survey team makes use of data products from the Near-Earth Object Wide-field Infrared Survey Explorer (NEOWISE), which is a project of the Jet Propulsion Laboratory/California Institute of Technology. NEOWISE is funded by the National Aeronautics and Space Administration.

The Legacy Surveys imaging of the DESI footprint is supported by the Director, Office of Science, Office of High Energy Physics of the U.S. Department of Energy under Contract No. DE-AC02-05CH1123, by the National Energy Research Scientific Computing Center, a DOE Office of Science User Facility under the same contract; and by the U.S. National Science Foundation, Division of Astronomical Sciences under Contract No. AST-0950945 to NOAO.

Based on observations obtained with MegaPrime/MegaCam, a joint project of CFHT and CEA/DAPNIA, at the Canada‑France‑Hawaii Telescope (CFHT), which is operated by the National Research Council (NRC) of Canada, the Institut National des Sciences de l’Univers of the Centre National de la Recherche Scientifique (CNRS) of France, and the University of Hawaii. The observations at the CFHT were performed with care and respect from the summit of Maunakea, which is a significant cultural and historic site.

The MeerLICHT telescope is designed, installed and operated by a
consortium including Radboud University, the University of Cape Town,
the University of Oxford, the University of Manchester, the University
of Amsterdam, and the South African Astronomical Observatory, a part of
the South African National Science Foundation.

\bibliographystyle{aa}
\bibliography{bibliography.bib} 
% Entries are in the bibliography.bib file

\appendix
\section{BGRem Hyper-Parameters: Normalization Factor and Diffusion Steps}\label{sec:app-diff-steps}
\begin{figure}
  \centering
  \subfloat{\includegraphics[scale=0.37]{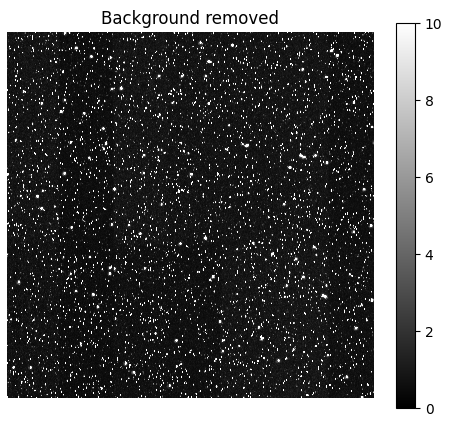}}
  \subfloat{\includegraphics[scale=0.37]{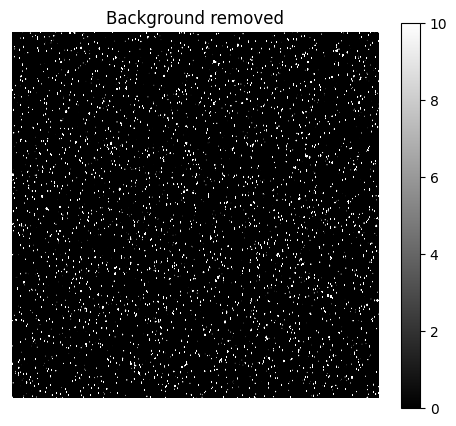}}
  \subfloat{\includegraphics[scale=0.37]{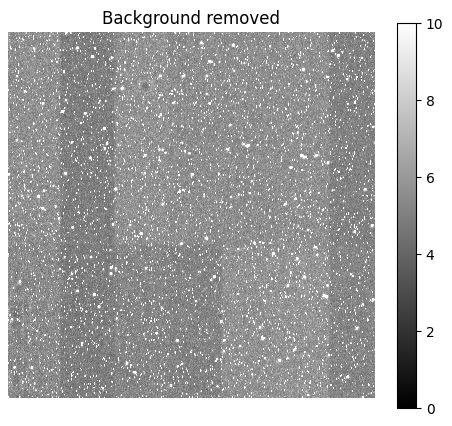}}
  \caption{The effect of a good normalization factor (left), a too low normalisation factor (middle) and a too high normalization factor (right) on the output of the model. } %\suvo{Rodney, I suggest to put this on the appendix or add it on a notebook that you will upload on GitHub}
  \label{fig:normalisation-factor}
\end{figure}

To get the best results with BGRem, here we discuss the choices of the two hyperparameters that might affect the performance. First, the normalization factor, which determines how well the background gets removed. Since the noise in the training data is likely different from the one encountered in most real images, we have to make them match as well as possible. This is done by tuning the normalization factor in such a way that the noise has a standard deviation of around 1. The mean of the noise isn't important since this is removed in one of the pre-processing steps. Setting the normalization factor too low will result in the model recognizing faint stars as background and removing them, while setting it too high will result in the model seeing the background as sources and not removing it. This is shown in Figure \ref{fig:normalisation-factor}.

BGRem has a built-in tool that can estimate the normalization factor by calculating the standard deviation $\sigma$ of the pixel values. Stars heavily influence this, so it ignores the 10\% brightest pixels. It also ignores all pixels with a value of 0 if there are any, so it doesn't take into account empty parts of images. The normalization factor is then simply $\frac{1}{\sigma}$. This works best for an image with a uniform background and might need tuning for images with large gradients in the background.

\begin{figure}
\centering
\includegraphics[width=0.9\columnwidth]{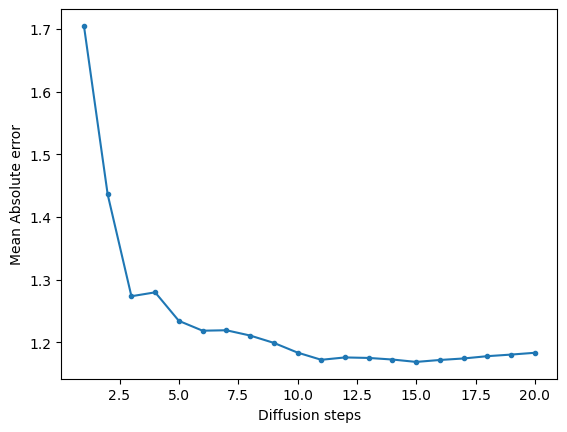}
\caption{The mean absolute error of BGRem on the validation data as a function of the number of diffusion steps.}
\label{fig:diffusion-steps-mae}
\end{figure}

The second parameter of BGRem is the number of diffusion steps. This determines how many steps the diffusion model removes the background. The accuracy increases with the increasing number of steps, but the computation time also increases linearly. Figure \ref{fig:diffusion-steps-mae} shows the mean absolute error of the model as a function of the number of diffusion steps. This was tested on the validation data.

Here we see that even with only one diffusion step, the model is very accurate. However, with more diffusion steps, the model is less likely to introduce image artefacts, making the performance even better. The optimal performance is at 15 diffusion steps, while 3, 6 and 11 are local optima, making these four choices preferable over other options.

%\section{Photon Counts of Individual Sources after BGREem}\label{sec:App2}
%A key concern for any model, in this case BGRem, is whether it alters the sources of interest after performing the background removal task. 
%
%Figure \ref{fig:fluxes} shows that BGRem is more consistent at predicting the total flux of different types of sources (including stars, galaxies, and elongated galaxies) than SExtractor across all signal-to-noise ratios (SNRs), with a smaller spread around the ideal line. BGRem also has the advantage of never producing negative values. We also highlight that for very bright sources (high SNR), our predicted pixel counts obtained with BGRem on the image cutouts follow $F_{\text{true}}$ values, showing that BGRem does not produce any artefacts as part of the background removal task.  
%
%At low SNRs, there is a bias toward lower fluxes, similar to SExtractor, but to a lesser extent. This occurs because BGRem sometimes identifies very faint sources as background and removes them. .   

\section{Effect of Training with MSE Loss}\label{sec:app-mse}

\begin{figure}
	\includegraphics[width=0.95\columnwidth]{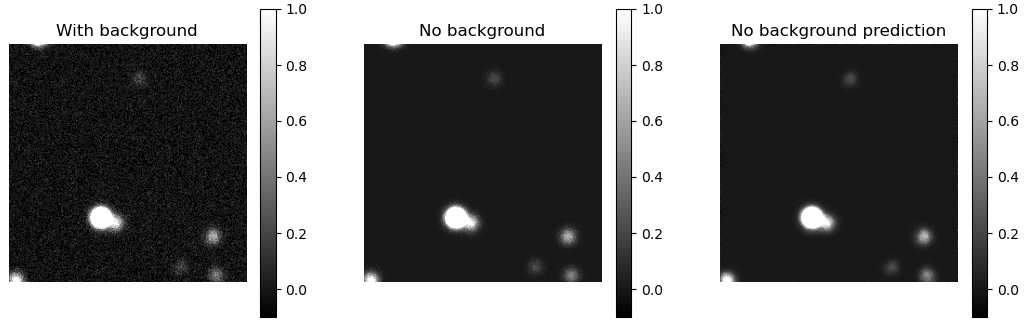}
	\includegraphics[width=0.95\columnwidth]{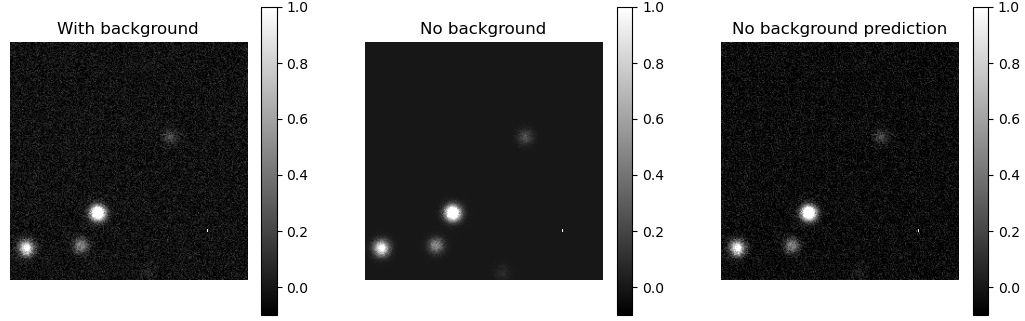}
	\caption{\rev{Example of training BGRem with MAE loss for a random MeerLICHT image (top row) compared with MSE loss (bottom row). Visually inspecting the final column where we obtain the `cleaned' image with BGRem highlights that training with MSE loss fails to remove background properly.}}
	\label{fig:bgrem-optical-mse}
\end{figure}

\begin{figure*}
	\centering
	\subfloat{{\includegraphics[width=0.3\linewidth]{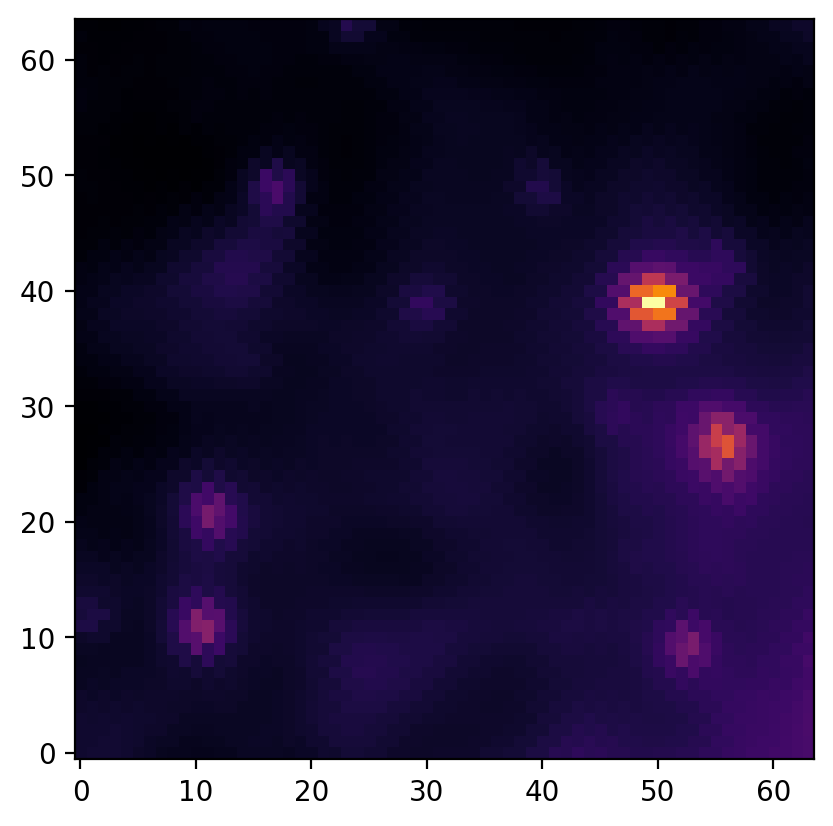} }}%
	\qquad
	\subfloat{{\includegraphics[width=0.3\linewidth]{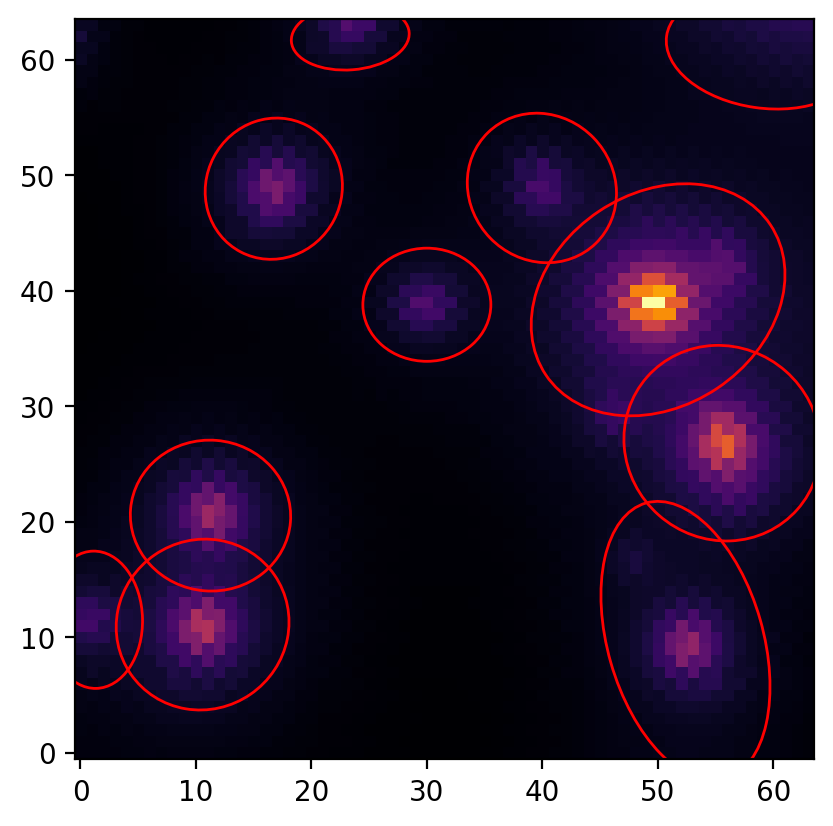} }}%
	\qquad
	\subfloat{{\includegraphics[width=0.3\linewidth]{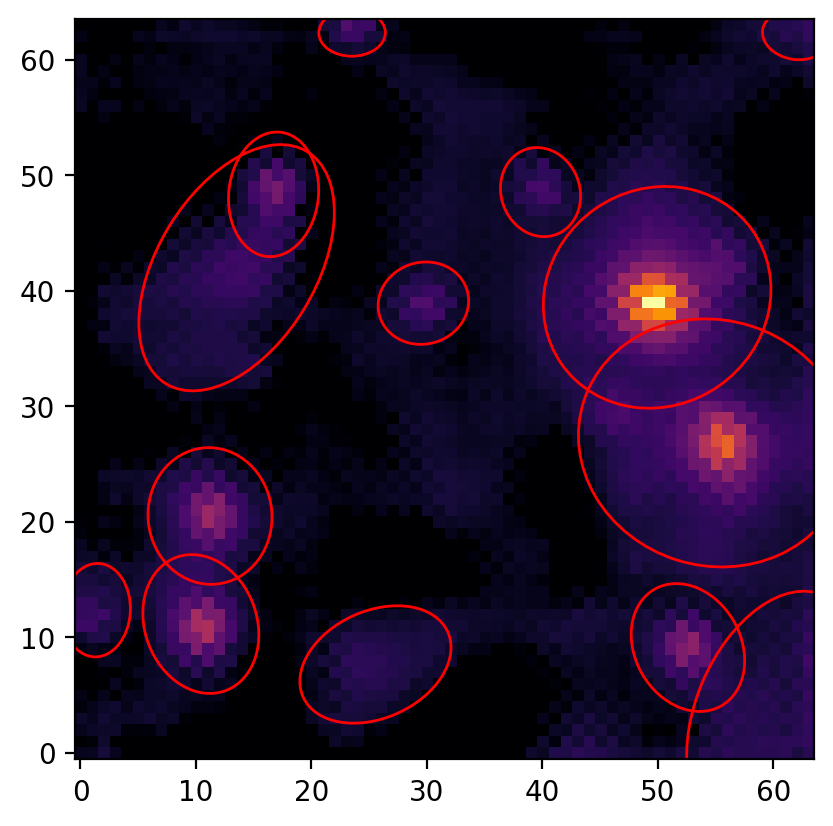} }}%
	
	\setcounter{subfigure}{0} % Reset subfigure counter
	
	\subfloat[\centering with Background]{{\includegraphics[width=0.3\linewidth]{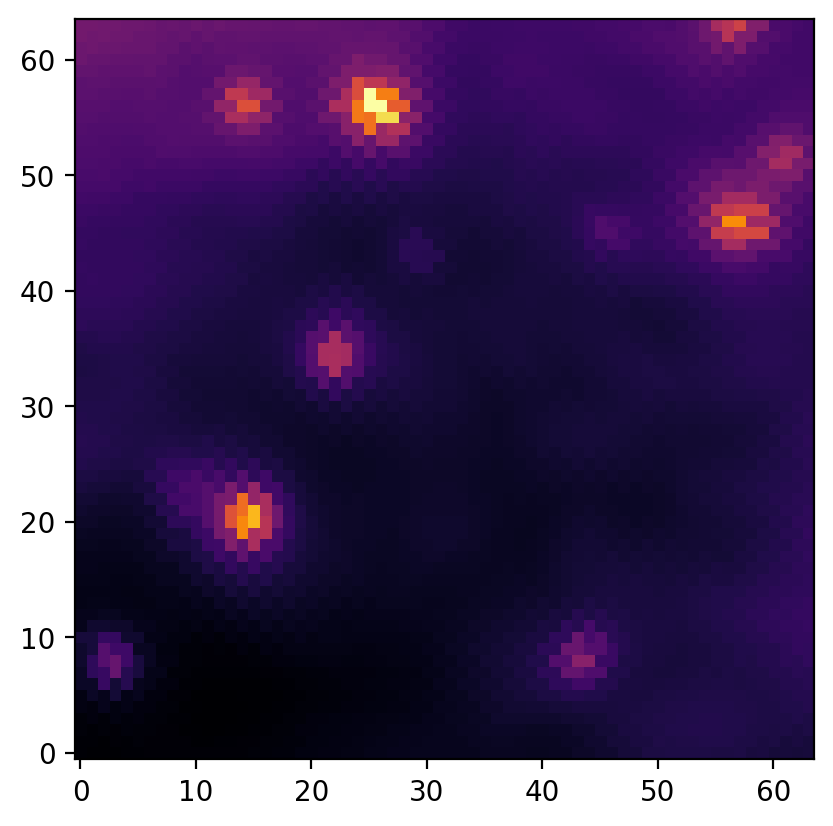} }}%
	\qquad
	\subfloat[\centering Source Only]{{\includegraphics[width=0.3\linewidth]{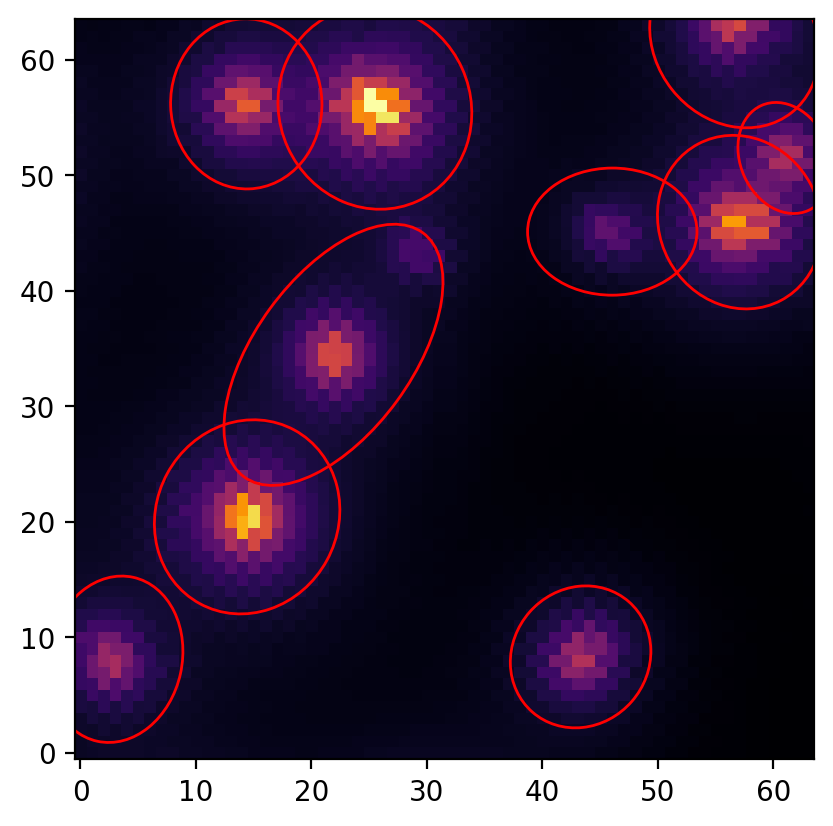} }}%
	\qquad
	\subfloat[\centering with BGRem]{{\includegraphics[width=0.3\linewidth]{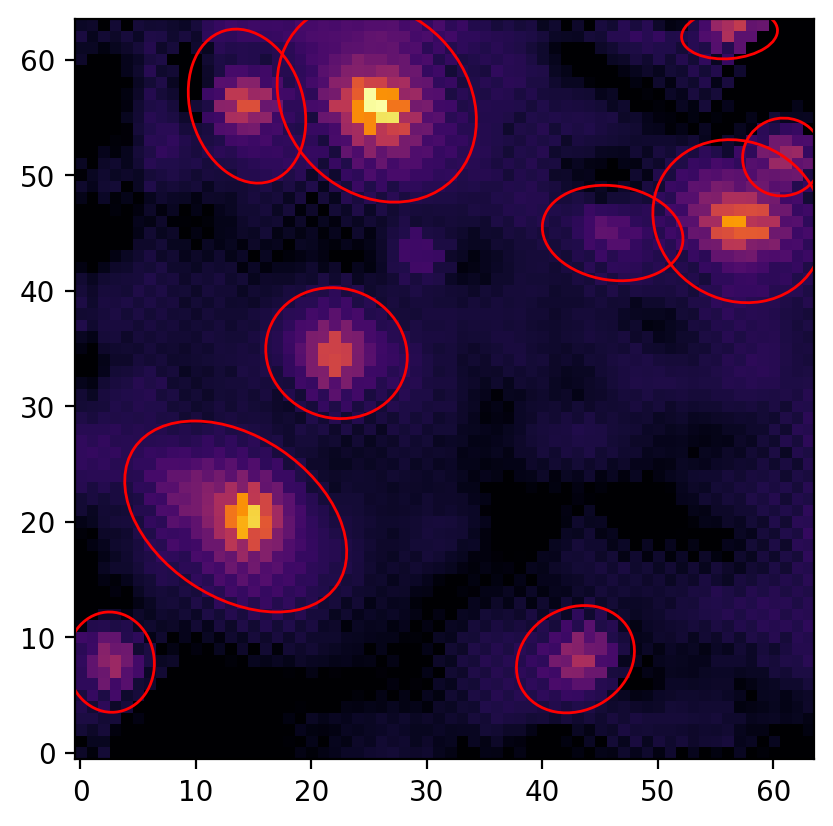} }}%
	\caption{\rev{Examples of BGRem for the same simulated gamma-ray sky patches as in Figure \ref{fig:example-bgrem-gamma}, but now obtained with MSE loss. It fails to properly remove the background noise and produces artifacts, highlighting MSE loss being less effective than MAE.}}%
	\label{fig:example-bgrem-gamma-wMSE}%
\end{figure*}

\begin{figure}
	\includegraphics[width=0.9\columnwidth]{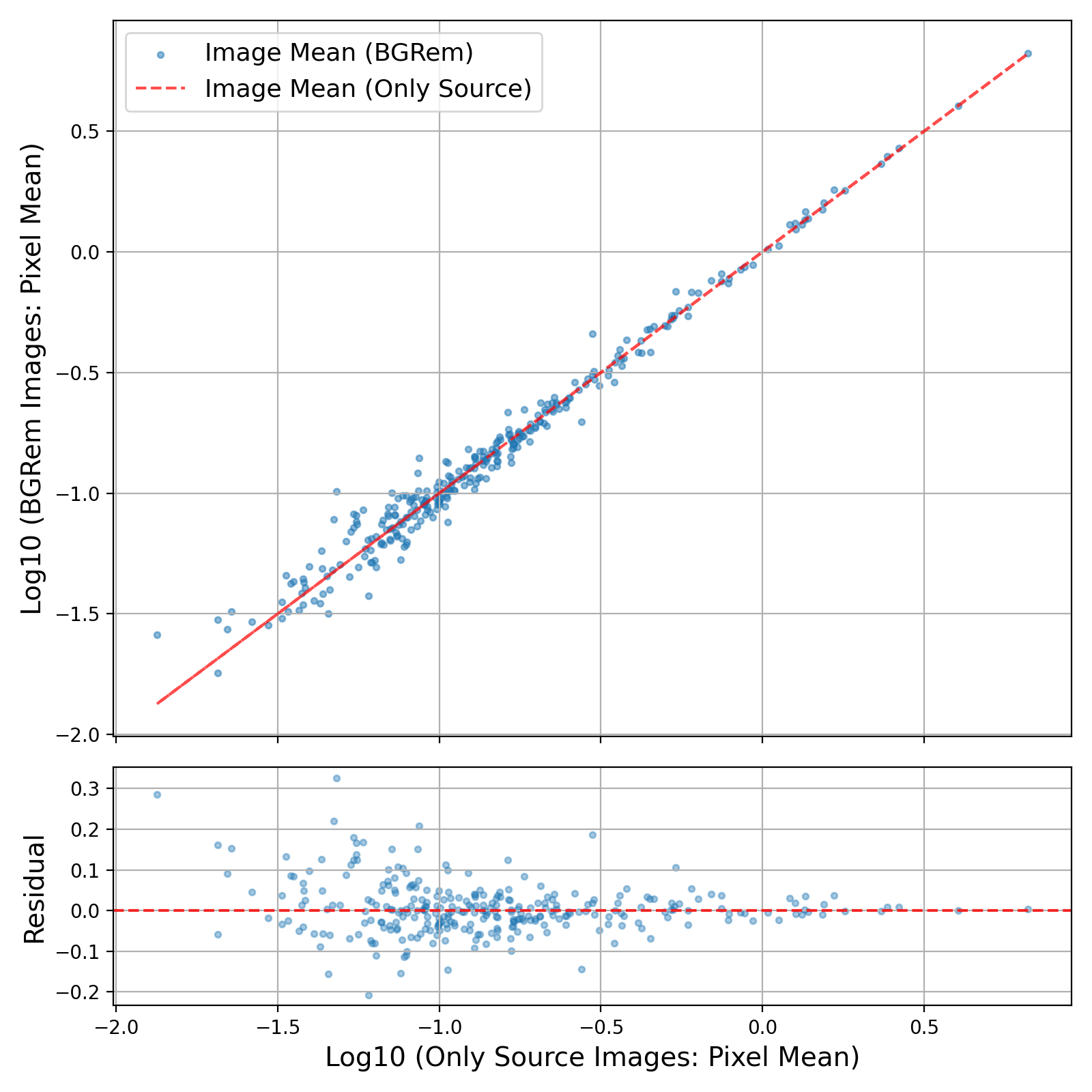}
	\caption{\rev{Same as Figure \ref{fig:bgrem-gamma-pix-mean} but now with MSE loss.}}
	\label{fig:bgrem-gamma-pix-mean-MSE}
\end{figure}

\rev{While in our work to train BGRem we have used MAE loss, in the original DDPM work, L2/MSE loss was introduced \citep{jon-ho-diffusion}. First, we checked from reconstructed image quality inspection that MSE loss fails to properly remove the background. This can be visually seen for reference images in optical dataset trained with MAE and MSE loss (top and bottom row of Figure \ref{fig:bgrem-optical-mse} respectively). The same can be seen for the $\gamma$-ray images as shown in Figure \ref{fig:example-bgrem-gamma-wMSE} and compared to the results presented before in Figure \ref{fig:example-bgrem-gamma}. Also, we quantify by plotting the mean pixel values of the denoised images with BGRem against the mean pixel values of the source-only images in Figure \ref{fig:bgrem-gamma-pix-mean-MSE}, which can be compared to the Figure \ref{fig:bgrem-gamma-pix-mean} presented before obtained with MAE loss. While for the brighter patches, even with MSE loss BGRem performance is similar to that of obtained with MAE loss, for the fainter patches the performance is significantly worse, which can be seen from the residuals.} 

\rev{In the original DDPM work, during the forward process the image is completely destroyed and usually we have a Gaussian noise with zero mean and unit variance $\epsilon \sim (\mathcal{N}(0, \mathbb{I}))$. Minimizing the MSE loss is a natural consequence of maximum likelihood under Gaussian Noise. For astronomical images with backgrounds (noise) we may have bright structures, localized excess, occasional extreme flares and these can be grouped into outliers; MAE penalizes error linearly and is less sensitive to outliers. While MAE performs better according to our tests, for our future work, we intend to test BGRem with Huber loss \citep{huber-loss} where it can seamlessly move from MSE to MAE depending on the small to large error values.}

%\begin{figure}
%	\includegraphics[width=0.9\columnwidth]{./L2loss.png}
%\end{figure}

\end{document}